\documentclass[12pt]{article} 

\usepackage{graphicx}

\def\1{{\mathchoice {1\mskip-4mu\mathrm l}
{1\mskip-4mu\mathrm l} {1\mskip-4.5mu\mathrm l}
{1\mskip-5mu\mathrm l}}}
\usepackage{amsmath}
\usepackage{amssymb}
\newtheorem{Lemma}{Lemma}[section]
\newcommand{\Z}{\mathbb{Z}}
\newcommand{\e}{\varepsilon}
\newtheorem{defi}{Definition}[section]
\newtheorem{propo}{Proposition}[section]
\newtheorem{Theorem}{Theorem}[section]

\title{Gibbsianness versus Non-Gibbsianness \\
of time-evolved planar rotor models}

\author{
{\normalsize A.C.D. van Enter }\\
{\normalsize Intitute for Mathematics and Computing Science,} \\
{\normalsize University of Groningen},\\
{\normalsize Nijenborgh 9, }\\
{\normalsize 9747AG,Groningen},\\
{\normalsize the Netherlands}\\
\\[3 mm]
{\normalsize W.M. Ruszel}\\
{\normalsize Intitute for Mathematics and Computing Science,} \\
{\normalsize University of Groningen},\\
{\normalsize Nijenborgh 9, }\\
{\normalsize 9747AG,Groningen},\\
{\normalsize the Netherlands}\\
{\normalsize and} \\
{\normalsize Center for Theoretical Physics,} \\
{\normalsize University of Groningen},\\
{\normalsize Nijenborgh 4, }\\
{\normalsize 9747AG,Groningen},\\
{\normalsize the Netherlands}}

\begin{document}
\maketitle

\newpage
\begin{abstract}
We study the Gibbsian character of time-evolved planar rotor systems on $\mathbb{Z}^d$, $d \geq 2$, in the transient regime, evolving with stochastic dynamics and starting from an initial Gibbs measure $\nu$. We model the system by interacting Brownian diffusions $X = (X_i(t))_{t \geq 0, i \in \mathbb{Z}^d}$ moving on circles. We prove that for small times $t$ and arbitrary initial Gibbs measures $\nu$, or for long times and both high- or infinite-temperature initial measure and dynamics, the evolved measure $\nu^t$ stays Gibbsian. Furthermore, we show that for a low-temperature initial measure $\nu$ evolving under infinite-temperature dynamics there is a time interval $(t_0,t_1)$ such that $\nu^t$ fails to be Gibbsian in $d=2$.
\end{abstract}

\section{Introduction}
During a stochastic evolution of Gibbs measures for spin systems various things can happen. In \cite{vEntFerHolRed02} the authors considered Ising-spin systems starting from an initial Gibbs measure $\nu$ and evolving under a spin-flip dynamics (high-temperature Glauber dynamics) towards a reversible Gibbs measure $\mu \neq \nu$, both having a finite-range interaction. They show that in this transient regime the evolved measure $\nu^t = \nu S(t)$ stays Gibbsian if either $t$ is small, or if both $\nu$ and $\mu$ are at high or infinite temperature. It can lose the Gibbs property after some time $t$ if the initial measure is at low temperature and zero or small magnetic field and $\mu$ is at high or infinite temperature. The Gibbs measure property is also shown to be recovered after some time if the initial measure $\nu$ has a non-zero magnetic field.
In some sense the initial external field will be compensated by a
dynamic field, causing a hidden phase transition which makes the evolved
measure non-Gibbs. For large enough time when the dynamic field is too weak to do this, there is re-entrance into the Gibbsian regime again. Thus in that case the evolved measure is Gibbsian if the time is either small or large enough. Le Ny and Redig generalized in \cite{LeNRed02} the result for small times $t$ to more general dynamics. They consider spin systems with spins in $\lbrace 0,1 \rbrace^{\mathbb{Z}^d}$ and prove that Gibbsianness stays conserved when the system evolves under a more general local stochastic dynamics such as Kawasaki or mixtures of Glauber and Kawasaki dynamics.
\newline
The case of an infinite-temperature dynamics leads to the question of
whether an effective temperature can be defined \cite{OliPet06} in the transient
regime, or, in a different interpretation, how reliable noisy observations
are, see e.g. \cite{You89, PryBru95} or \cite{Tan02, SaiNis02}. The last type of questions lead to the study of
so-called ``Hidden Markov Fields''.
\newline
Another Gibbsian question of interest for discrete spins with discrete-time dynamics is the study of PCA's (probabilistic cellular automata) in $d$ dimensions. Their stationary measures are projections of space-time Gibbs measures for a Hamiltonian obtained by the PCA in $d+1$ dimensions in the stationary case. In \cite{GolKuiLebMae89} the authors prove the converse direction that all translation invariant or periodic stationary measures correspond to Gibbs measures on space-time histories for the PCA on a space $S^{\mathbb{Z}^{d+1}}$.

\medskip

What is known about the situation for continuous spins?
Deuschel in \cite{Deu87} and later Roelly, Zessin and coauthors investigated in \cite{MinRoeZes00} and \cite{CatRoeZes96} Gibbs measures of interacting diffusions indexed by lattice sites on $\mathbb{Z}^d$. First Deuschel described the law $Q^{\nu}$ of the entire infinite-dimensional unbounded process $X=((X_i(t))_{0 \leq t \leq 1, i \in \mathbb{Z}^d}$ as a space-time Gibbs measure with state space $C([0,1])^{\Z^d}$ when the initial distribution $\nu$ is Gibbsian. Then Roelly, Zessin and coauthors showed a bijection between the set of initial Gibbs measures associated to an initial interaction on $\mathbb{R}^{\Z^d}$ and the set of Gibbs measures on the path space describing the full dynamics, see \cite{CatRoeZes96}.
\newline
In \cite{DerRoe05} Dereudre and Roelly considered a problem which is close to the one we study, namely the Gibbsianness in the transient regime for the time evolution of unbounded continuous spins under high-temperature dynamics with bounded interactions. They start with the Gibbs representation of $Q^{\nu}$ on the path level and look at the projections at time $t$ of the law $Q^{\nu}$. Then they prove that if the initial measure obeys a strong uniqueness condition, then either for small times or if the dynamical interaction is weak enough, Gibbsianness propagates for bounded initial and dynamical interaction. That means that the time-evolved measure is again a Gibbs measure for continuous spins and an absolutely summable interaction. Another case of unbounded continuous spins was treated by K\"ulske and Redig in \cite{KueRed06} where the authors also consider the time evolution of continuous unbounded spin models under infinite-temperature dynamics and {\em unbounded} interaction (compared to Roelly and Dereudre who consider a bounded one). They prove that (similar to the result of \cite{vEntFerHolRed02} for discrete spins) for continuous unbounded spins the time-evolved measure stays Gibbsian for small times but loses this property for $t$ large if the initial measure is a low-temperature measure. In contrast to the discrete-spin situation the Gibbsian property cannot be recovered again.
The initial Gibbs measure is Gibbsian for a Hamiltonian with a quadratic nearest neighbour interaction and an a priori single-site double well potential that has a specific form. This particular form of the interaction term allows an explicit analysis. Their definition of a Gibbsian measure is weaker than the DLR definition imposing an absolutely summability condition on the interaction like \cite{DerRoe05}. In the case of unbounded spins it seems more natural to weaken the assumptions on the interaction since absolutely summability does not
even allow Gaussian fields.

\bigskip

In this paper we consider continuous but compact spins living on a circle $\mathbb{S}^1$. We investigate the Gibbsian property of the system first for small times $t$ and arbitrary initial Gibbs measures $\nu$ under different dynamics and then for small inverse initial temperature and arbitrary times. The purpose of the third section will be to show that the time-evolved measure $\nu^t$ is Gibbsian for small $t$ under infinite- or high-temperature dynamics and arbitrary-temperature initial Gibbs measure, and for any $t$ for high- or infinite-temperature dynamics starting in a high- or infinite-temperature initial measure. We follow and in our case can simplify the approach of \cite{DerRoe05}. Like in their approach we will get the time-evolved measure from projecting the path-space measure $Q^{\nu}$ at time $t$. In the fourth section we prove that the Gibbs property will be lost after some time $t$ for a low-temperature initial measure and infinite-temperature dynamics, somewhat similarly to the situation in \cite{KueRed06} for the classical nearest neighbour plane rotor in $d=2$ with zero external initial field.
This may seem slightly surprising, since the (translation-invariant)
Gibbs measure is unique at all temperatures (and in all homogeneous external
fields). In our opinion the advantage of working with compact spins, next to the technical simplifications, is that the notion of Gibbsianness for a ``reasonable'' class of interaction is much less ambiguous. See e.g. the discussion in \cite{vEnFerSok93}. Closely related results have been obtained by K\"ulske and Opoku \cite{KueOpo07} via a different approach using Dobrushin uniqueness techniques. The authors investigate short-time behaviour of spins moving on $\mathbb{S}^{N-1}$ and evolving with infinite-temperature dynamics.

\section{General framework}

We introduce some definitions and notations. The state space of one continuous spin is the circle, $\mathbb{S}^1$. We identify the circle with the interval $[0,2\pi)$ where $0$ and $2\pi$ are considered to be the same points. Thus the configuration space $\Omega$ of all spins is isomorphic to $[0,2\pi)^{\Z^d}$. We endow $\Omega$ with the product topology and natural product measure $\nu_0(dx) = \bigotimes_{i \in \mathbb{Z}^d} \nu_0(d x_i)$. In our case we take $\nu_0(d x_i) = \frac{1}{2 \pi} dx_i$.
An interaction $\varphi$ is a collection of $\mathcal{F}_{\Lambda}$-measurable functions $\varphi_{\Lambda}$ from $([0,2\pi))^{\Lambda}$ to $\mathbb{R}$ where $\Lambda \subset \mathbb{Z}^d$ is finite.
$\mathcal{F}_{\Lambda}$ is the $\sigma$-algebra generated by the canonical projection on $[0,2\pi)^{\Lambda}$. The interaction $\varphi$ is said to be of \textbf{finite range} if there exists a $r > 0$ s.t. $diam(\Lambda) > r$ implies $\varphi_{\Lambda} \equiv 0$ and it is called \textbf{absolutely summable} if for all $i$, $\sum_{\Lambda \ni i} \parallel \varphi_{\Lambda} \parallel_{\infty} < \infty$.
We call $\mu$ a \textbf{Gibbs measure} associated to a reference measure $\nu_0$ and interaction $\varphi$ if the series $H_{\Lambda}^{\varphi}= \underset{\Lambda^{\prime} \cap \Lambda \neq \emptyset} \sum \varphi_{\Lambda^{\prime}} $ converges ($\varphi$ is absolutely summable) and $\mu$ satisfies the DLR equations for all $i$:

\begin{equation*}
\mu(dx_i \mid x_j, j \neq i) = \frac{1}{Z_i} \exp( - H_{i}^{\varphi}(x)) \nu_0( dx_i),
\end{equation*}

where $Z_i = \frac{1}{2\pi} \int_0^{2\pi} \exp( - H_{i}^{\varphi}(x)) dx$ is the partition function. We omit the inverse temperature $\beta$ here, because in this section we deal with arbitrary initial inverse temperature. The set of all Gibbs measures associated to $\varphi$ and $\nu_0$ is denoted by $\mathcal{G}( \varphi, \nu_0)$ (resp. $\mathcal{G}_{\beta}( \varphi, \nu_0)$ if we want to make the $\beta$ dependence clear). \newline

Furthermore we say an interaction $\varphi$ satisfies \textbf{a high-temperature Dobrushin condition} if it is absolutely summable and
\begin{equation}
\sup_{i \in \Z^d} \sum_{ \Lambda \ni i} (|\Lambda| -1) \sup_{x, y} |\varphi_{\Lambda}(x) - \varphi_{\Lambda}(y)| < 2. \label{Dobrushin}
\end{equation}
We will also refer to such an interaction as a \textit{high-temperature interaction} and use the fact that if an interaction satisfies the condition above it follows that $|\mathcal{G}(\varphi, \nu_0)| = 1$, see e.g. \cite{Geo88}. \newline

Now, instead of working with Gibbs measures on $[0,2\pi)^{\Z^d}$ we will first investigate Gibbs measures as space-time measures on the path space $\overset{\sim} \Omega = C(\mathbb{R}_+, [0,2\pi))^{\mathbb{Z}^d}$. In \cite{Deu87} Deuschel introduced and descibed infinite-dimen\-sional diffusions as Gibbs measures on the path space $C([0,1])^{\Z^d}$ when the initial distribution is Gibbsian. This approach was later generalized by $\cite{CatRoeZes96}$ who showed that there exists a one-to-one correspondence between the set of initial Gibbs measures and the set of path-space measures $Q^{\nu}$.
\\

More precisely, let $X=(X_i(t))_{t \geq 0, i\in \Z^d}$ be the solution of the following
stochastic differential equation (SDE)
\begin{eqnarray}
\begin{cases}
& d X_i(t) = -\nabla_i H_i^{\varphi}(X(t))dt + d B_i^{\odot}(t) , t > 0, i \in \mathbb{Z}^d \\
& X(0) \simeq \nu , t=0 \label{system1}
\end{cases}
\end{eqnarray}
where $\nu$ will be the initial Gibbs measure, $\nu \in \mathcal{G}(\overset{\sim} \varphi, \nu_0)$ associated to some initial interaction $\overset{\sim} \varphi$ which is supposed to be of finite range and at least $C^2$. $B^{\odot}=(B^{\odot}_i(t))_{t \geq 0, i \in \Z^d}$ denotes a collection of independent Brownian motions on the circle at each lattice site $i$. $\nabla_i$ is a differential operator on the circle at lattice site $i$ (equal to $\frac{d}{d x_i}$). In other words, the system at time 0 starts in a Gibbs distribution $\nu$ and evolves, according to a Brownian motion on the circle with some drift.
Basically $X_i(t)$ describes the position of the rotor spin at site $i$ at time $t$, it takes values in $X_i(t)(\omega) \in [0,2\pi)$ for every event $\omega$. \\
 
The interaction part of the stochastic dynamics of this process is hidden in the drift term of the SDE, namely in the gradient of the Hamiltonian $H^{\varphi}$. The underlying dynamical interaction $\varphi$ is also assumed to be of finite range and at least $C^2$ so that the Gibbs measures for it are transient and reversible. We will refer to different temperatures for the dynamics. What we mean by that is the following: Let the Gibbs measure $\nu$ converge towards a reversible Gibbs measure $\mu$. Then we say the dynamics is low, high or infinite temperature if the corresponding infinite-time measure $\mu$ is a low-, high- or infinite-temperature Gibbs measure. Since the spins are compact, this implies that the derivatives are automatically bounded. The assumptions on $\nu$ and $\varphi$ provide the existence of a strong Markovian solution $X$ of the above system $\eqref{system1}$. Let furthermore $Q^{\nu}$ denote the law of the solution $X$ on $\overset{\sim} \Omega$ with initial distribution $\nu$. \\

We first note that the initial Gibbs measure $\nu$ is a mixture of extremal
Gibbs measures from the set $\mathcal{G}(\overset{\sim} \varphi, \nu_0)$,
see also the representation theorem (Theorems 7.12 and 7.26 in \cite{Geo88}).
Note that in general the mixture of $Q^x$ under a measure $\eta$ is given by
$Q^{\eta} = \int Q^x \eta(dx)$. This means that if $\mu$ is an extremal element
of $\mathcal{G}(\varphi, \nu_0)$, then there exists a
$y \in [0,2\pi)^{\mathbb{Z}^d}$ such that

\begin{equation}
\mu = \underset{\Lambda \rightarrow \mathbb{Z}^d} \lim \mu_{\Lambda, y} \otimes \delta_{y_{\Lambda^c}} \label{extremal}
\end{equation}

where $\mu_{\Lambda, y} (dx) = \frac{1}{Z^y_{\Lambda}} \exp (- H^{\varphi}_{\Lambda, \Lambda^c}(x,y) ) \nu_0^{\otimes \Lambda} (dx_{\Lambda})$ is the finite-volume Gibbs measure with fixed boundary condition $y$.
\newline

Let first $\nu \in \mathcal{G}(\overset{\sim} \varphi, \nu_0)$ be extremal and let $(\nu_{\Lambda,y})_{\Lambda}$ denote the approximating sequence as in $\eqref{extremal}$. Then we have
\begin{equation}
\nu_{\Lambda,y}(dx_{\Lambda})= \frac{1}{Z^y_{\Lambda}} \exp(- \overset{\sim}H^{\overset{\sim} \varphi}_{\Lambda,\Lambda^c}(x,y)) \nu_0^{\otimes \Lambda} (dx_{\Lambda}). \label{repreNu}
\end{equation}
The tilde in $\overset{\sim}H^{\overset{\sim} \varphi}$ and $\overset{\sim}\varphi$ will always refer to the Hamiltonian, resp. the interaction belonging to the initial distribution $\nu$. Let us denote the set of all Gibbs measures which are obtained as weak limit points of finite-volume measure by $\mathcal{G}_0(\overset{\sim} \varphi, \nu_0)$ and remark that $\mathcal{G}_0(\overset{\sim} \varphi, \nu_0) \subset \mathcal{G}(\overset{\sim} \varphi, \nu_0)$. \\

We define the $i$-decoupled initial measure $\nu^i_{\Lambda,y}$ as
\begin{eqnarray*}
\nu^{i}_{\Lambda,y}(dx_{\Lambda}) & = & \frac{1}{Z^y_{\Lambda,i} }\exp(- \overset{\sim}H^{\overset{\sim} \varphi}_{\Lambda \setminus i, \Lambda^c}(x,y)) \frac{1}{(2 \pi)^{\mid \Lambda \mid - 1}} dx_{\Lambda \setminus i} \frac{1}{2\pi} dx_i \\
& = & Z^y_{\Lambda,i} \exp(\overset{\sim}H_i^{\overset{\sim}\varphi}(x_{\Lambda}, y_{\Lambda^c})) \nu_{\Lambda,y}(d x_{\Lambda}). 
\end{eqnarray*}
The decoupled Hamiltonian leaves out of the summation all sets containg site $i$.
We notice that $\nu^i_{\Lambda,y} \bigotimes \delta_{y_{\Lambda^c}}$ converges in $\Lambda$ towards a measure $\nu^i$ on $[0,2\pi)^{\mathbb{Z}^d}$ with the following properties
\begin{itemize}
\item $\nu << \nu^i$ (absolutely continuous) and
\item $\nu(dx)=\frac{1}{Z_i}\exp(-\overset{\sim}H^{\overset{\sim}\varphi}_i(x)) \nu^i(dx)$.
\end{itemize}

Let us write down the finite-dimensional $i$-decoupled dynamics:
\begin{eqnarray*}
\begin{cases}
& d X_j(t) = -\nabla_j H_j^{\varphi}(X(t))dt + d B_j^{\odot}(t), j \in \Lambda, j \neq i, t>0 \\
& d X_i (t) = d B_i^{\odot}(t), t>0
\end{cases}
\end{eqnarray*}
that is at lattice site $i$ we put a Brownian motion $B^{\odot}_i$ and the dynamics on the other lattice sites depend on all spins $j$ which are not equal to $i$. In the ``decoupled regime" we start with an $i$-decoupled measure and evolve with $i$-decoupled dynamics. We denote by $Q^{x_{\Lambda}}_{\Lambda}$ (resp. $Q^{x_{\Lambda},i}_{\Lambda}$) the law of the solution of the (resp. $i$-decoupled) system of the finite system on $\Lambda$ with fixed initial condition $x_{\Lambda}$. \\

Moreover, the measures
\begin{equation}
\underset{\Lambda \rightarrow \Z^d} \lim \nu_{\Lambda, y}^t = \nu^t \text{ resp. } \underset{\Lambda \rightarrow \Z^d} \lim \nu_{\Lambda, y}^{t,i}= \nu^{t,i} \label{weakconv}
\end{equation}
converge in a weak sense. Here also no difficulties arise because of compactness of the spins.
\medskip

If $t$ is large we divide, like \cite{DerRoe05}, the dynamical interaction into two parts, $U + \beta \varphi$. $U$ is the self-interaction, which is a single-site term, and $\varphi$ is the interaction proper, while $\beta$ will serve as a small parameter. The term $U$ can also be viewed as the single-site drift term of the Brownian motions moving on the circle. The system will be defined as follows
\begin{eqnarray}
\begin{cases}
& d X_i (t) = - \frac{1}{2} U^{\prime}(X_i(t))dt - \frac{\beta}{2}\nabla_i H^{\varphi}_i(X(t)) dt + d B^{\odot}_i(t), \text{ } i \in \mathbb{Z}^d, t > 0 \label{sys1}\\
& X(0) \simeq \nu , \text{ } t=0
\end{cases}
\end{eqnarray}
and for $\beta = 0$
\begin{eqnarray}
\begin{cases}
& d X_i (t) = - \frac{1}{2} U^{\prime}(X_i(t))dt + d B^{\odot}_i(t) , \text{ } i \in \mathbb{Z}^d, t > 0 \label{sys2} \\
& X(0) \simeq \nu , \text{ } t=0.
\end{cases}
\end{eqnarray}
For long times the role of the inverse temperature $\beta$ will be important, so we include it into the definition of the process. We denote in the same spirit as before $Q^{\nu}_{\beta}$ the law of the solution of $\eqref{sys1}$ resp. $Q^{\nu}_{0}$ the law of the solution of $\eqref{sys2}$. We call
$S$, the space $[0,2\pi)^{\Z^d \times \lbrace 0, 1 \rbrace}$, the bi-space. It will represent a double-layer system, to treat the distribution at time 0 and at the $t$ simultaneously. The joint distribution on the bi-space will be denoted by $\textbf{Q}^{\nu}_{\beta} = Q^{\nu}_{\beta} \circ (X(0),X(t))^{-1}$. So we project the path-space measure $Q^{\nu}$ at time $0$ and $t$.
\newline

We will use the following proposition to detect Gibbsianness of the evolved measure $\nu^t$.
\begin{propo}\label{equivPropo}
The following statements are equivalent:
\begin{enumerate}
\item $\mu$ is a Gibbs measure.
\item For all configurations $\eta$ and finite $\Lambda \subset \mathbb{Z}^d$ the the measure $\mu$ admits a continuous and strictly positive version of its conditional probabilities.
\item $\mu$ admits a continuous version of the Radon-Nikod\'ym derivatives $\frac{d \mu^i}{d \mu}$ for all $i \in \mathbb{Z}^d$ in the product topology.
\end{enumerate}
\end{propo}
A proof can be found in e.g. \cite{Sul73} and for continuous spins \cite{Geo88}. To determine if a measure is Gibbs or not we will mainly use the third item. For the failure of Gibbsianness we will use the necessary and sufficient condition of finding a point of essential discontinuity of (every version of) the conditional probabilities of $\mu$, i.e. a so-called "bad configuration".
It is defined as follows
\begin{defi}\label{badconfig}
A configuration $\zeta$ is called \textbf{bad} for a probability measure $\mu$ if there exists a $\e > 0$ and $i \in \Z^d$ such that for all $\Lambda$ there exists $\Gamma \supset \Lambda$ and configurations $\xi$, $\eta$ such that
\begin{equation*}
|\mu_{\Gamma}(X_i | \zeta_{\Lambda \setminus \lbrace i \rbrace}\eta_{\Gamma \setminus \Lambda}) - \mu_{\Gamma}(X_i | \zeta_{\Lambda \setminus \lbrace i \rbrace}\xi_{\Gamma \setminus \Lambda}) | > \e.
\end{equation*}
\end{defi}

\section{Conservation of Gibbsianness}

\subsection{Small times}

Let the dynamical interaction $\varphi$ be of finite range and for every $\Lambda \subset \Z^d$ finite, let $\varphi_{\Lambda}$ be $C^2([0,2\pi)^{\Lambda})$. Denote by $H_i^{\varphi}$ the associated Hamiltonian function ($H_i^{\varphi} = \sum_{A: A \cap \lbrace i \rbrace \neq \emptyset} \varphi_A $). We consider now the process $X=(X_i(t))_{t \geq 0, i \in \mathbb{Z}^{d}}$ defined by
\begin{eqnarray}
\begin{cases}
& d X_i(t) = -\nabla_i H_i^{\varphi}(X(t))dt + d B_i^{\odot}(t), i \in \Z^d, t > 0 \label{system3} \\
& X(0) \simeq \nu , t=0.
\end{cases}
\end{eqnarray}
where $\nu \in \mathcal{G}(\overset{\sim} \varphi, \frac{1}{2 \pi}dx)$ with $\overset{\sim} \varphi$ of also finite range and $C^2$. In particular since the spins are compact, $\overset{\sim} \varphi$ and $\varphi$ are also Lipschitz continuous. Conservation of Gibbsianness for small times does not need any constraints on either the initial or the dynamical temperature, so we will not specify them. Then we have the following theorem.
\begin{Theorem}
Let $Q^{\nu}$ be the law of the solution $X$ of the system $\eqref{system3}$. Let
$\nu \in \mathcal{G}(\overset{\sim}\varphi, \frac{1}{2\pi}dx)$ and let $\varphi$ satisfy the conditions above. Then there exists a time $t_0(\varphi, \overset{\sim} \varphi) > 0$ s.t. for all $t \leq t_0$ there exists an absolutely summable interaction $\varphi^t$ for which $\lbrace \nu^t=Q^{\nu}\circ X(t)^{-1}: \nu \in \mathcal{G}(\overset{\sim} \varphi, \frac{1}{2 \pi} dx) \rbrace \subset \mathcal{G}(\varphi^t, \frac{1}{2\pi}dx)$ is a Gibbs measure. The evolved interaction $\varphi^t$ depends only on the time $t$, the initial interaction $\overset{\sim} \varphi$ and the dynamical interaction $\varphi$.
\end{Theorem}
\textbf{Proof:}\\
The proof follows basically that of \cite{DerRoe05}. Some results will follow easier since we are dealing with compact spins. The first part of the argument is essentially the same. The scheme will be as follows. \\ 
To identify $\nu^t$ as being a Gibbs measure for small times we want to use proposition $\ref{equivPropo}$. Therefore we need to know how the Radon-Nikod\'ym derivative, $\frac{d \nu^t}{d \nu^{t,i}}(x)$, looks like for every $t$ and $i$. (We will prove the theorem for $\nu$ being an extremal Gibbs measure for convenience.) Therefore we first compute
$\frac{d \nu^t_{\Lambda, y}}{d \nu^{t,i}_{\Lambda, y}}(x_{\Lambda})$, the Radon-Nikod\'ym derivative of the projected law of the finite-dimensional system in $\Lambda \subset \Z^d$ at time $t$ and some boundary condition $y_{\Lambda^c}$ outside $\Lambda$. Then using cluster expansion techniques and the weak convergence $\eqref{weakconv}$ we will be able to demonstrate that for $t$ small, this derivative is continuous and behaves nicely. Next we can
use the Kotecky-Preiss criterion, see \cite{KotPre86}, to deduce that this expansion converges, uniformly in $\Lambda$, $x$ and the boundary condition $y$. For $\Lambda$ going to infinity this RN-derivative approaches a continuous function. That will be enough to deduce the existence of an absolute summable interaction $\varphi^t$ for which $\nu^t$ is Gibbs if $t$ is small.

\medskip

In the following we fix a boundary condition $y_{\Lambda^c}$ and compare the time-evolved finite-dimensional distribution $\nu^t_{\Lambda, y}$ with the $i$-decoupled one $\nu^{t, i}_{\Lambda, y}$.
By some algebraic manipulation we get for the Radon-Nikod\'ym derivative
 $\frac{d \nu^t_{\Lambda, y}}{d \nu^{t,i}_{\Lambda, y}}(x_{\Lambda})$ that it has the following form.
\begin{Lemma}\label{lemmaRN}
Let $t > 0$, $\Lambda \subset \Z^d$ and $i \in \Lambda$, then we have
\begin{equation*}
\frac{d \nu^t_{\Lambda, y}}{d \nu^{t,i}_{\Lambda, y}}(x_{\Lambda}) = \exp(- \overset{\sim}H_i^{\overset{\sim} \varphi}(x_{\Lambda},y_{\Lambda^c})) \frac{\mathbb{E}_{Q_{\Lambda}^{x_{\Lambda}}}(\exp(f_{\Lambda,y}(X_{\Lambda})(t) - f_{\Lambda,y}(x_{\Lambda})))}{\mathbb{E}_{Q_{\Lambda,i}^{x_{\Lambda}}}(\exp(f_{\Lambda\setminus i ,y}(X_{\Lambda})(t) - f_{\Lambda \setminus i,y}(x_{\Lambda})))} 
\end{equation*}
where
\begin{equation*}
f_{\Lambda,y}(x)= H^{\varphi}_{\Lambda, \emptyset}(x_{\Lambda})- \overset{\sim} H^{\overset{\sim} \varphi}_{\Lambda, \Lambda^c}(x,y).
\end{equation*}
\end{Lemma}
$\mathbb{E}_{Q_{\Lambda}^{x_{\Lambda}}}$ means the expectation with respect to the measure $Q_{\Lambda}^{x_{\Lambda}}$. Let us remark that for $t$ small $f_{\Lambda,y}(X_{\Lambda})(t) - f_{\Lambda,y}(x_{\Lambda})$ will turn out to be close to 0.
We fix the time $t$ and use Girsanov's theorem and write the measures $Q_{\Lambda}^{x_{\Lambda}}$ and the decoupled one $Q_{\Lambda, i}^{x_{\Lambda}}$ in terms of densities w.r.t. the product $\bigotimes_{j \in \Lambda} \rho^{\odot, x_j}$. The terms $\rho^{\odot, x_j}$ denote the ``Wiener measure on the circle" with initial condition $x_j$. In other words, we want to find a function $F^t_{\Lambda}$, such that
\begin{equation*}
Q^{x_{\Lambda}}_{\Lambda, y}(d X_{\Lambda}) = F^t_{\Lambda}(X_{\Lambda}) \bigotimes_{j \in \Lambda} \rho^{\odot, x_j}(d X_{j}).
\end{equation*}

Girsanov's theorem gives us the following form of $F^t_{\Lambda}$.
\begin{eqnarray*}
& & F^t_{\Lambda}(X_{\Lambda}) := \\
& & \exp \sum_{i \in \Lambda} \biggl( \int_0^t -\frac{1}{2} \biggl( \frac{d}{d x_i} H_{i, \Lambda}^{\varphi}(X_{\Lambda}(s)) \biggr ) dX_i(s) - \frac{1}{8} \int_0^t \biggl( \frac{d}{d x_i} H^{\varphi}_{i, \Lambda} \biggr)^2 (X_{\Lambda}(s)) d s \biggr ).
\end{eqnarray*}
and using It\^o's formula, we can write the function $F^t_{\Lambda}(X_{\Lambda})$ as
\begin{eqnarray*}
& & F^t_{\Lambda}(X_{\Lambda}) \\
& = & \exp \biggl ( -\frac{1}{2} H^{\varphi}_{\Lambda, \emptyset}(X_{\Lambda}(t)) + \frac{1}{2} H^{\varphi}_{\Lambda, \emptyset}(X_{\Lambda}(0)) \biggr ) \times \\
& & \times \exp \biggl( \sum_{i \in \Lambda} \int_0^t \biggl [ \frac{1}{4} \biggl (\frac{d^2}{d x_i^2} H^{\varphi}_{\Lambda, \emptyset} \biggr ) - \frac{1}{8} \biggl( \frac{d}{d x_i} H^{\varphi}_{\Lambda, \emptyset} \biggr)^2 \biggr ] (X_{\Lambda}(s)) ds \biggr ) \\
& =: & \exp \biggl (- \sum_{A \subset \Lambda} \Phi^{\varphi, t}_A (X_{\Lambda}) \biggr )
\end{eqnarray*}
with
\begin{equation*}
\Phi^{\varphi,t}_A (X_{\Lambda}) := \frac{1}{2} \varphi_A(X_{\Lambda}(t)) - \frac{1}{2} \varphi_A(X_{\Lambda}(0)) -
\end{equation*}
\begin{equation*}
\int_0^t \biggl ( \frac{1}{4} \sum_{j \in A} \frac{d^2}{ dx_j^2} \varphi_A(X_{\Lambda}(s)) - \frac{1}{8} \sum_{\substack{B \cup C = A \\ B\cap C \neq \emptyset }} \sum_{j \in B \cap C} \frac{d}{d x_j} \varphi_B(X_{\Lambda}(s)) \frac{d}{d x_j} \varphi_C (X_{\Lambda}(s)) \biggr ) ds. 
\end{equation*}
The collection $\Phi^{\varphi,t} = (\Phi^{\varphi,t}_A)_{A \subset \Z^d}$ forms an interaction potential on $\overset{\sim} \Omega$ on events that depend only on times between 0 and $t$.
Let $H^{\Phi^t}$ denote the Hamiltonian associated to $\Phi^{\varphi, t}$ then we can write finally
\begin{equation}
Q_{\Lambda, y}^{x_{\Lambda}} (d X_{\Lambda}) = \exp(- H^{\Phi^t}_{\Lambda,\emptyset}(X_{\Lambda})) \bigotimes_{j \in \Lambda} \rho^{\odot, x_j}(d X_j) \label{Girsanov}
\end{equation}
and analogously for the decoupled measure
\begin{equation*}
Q_{\Lambda, i, y}^{x_{\Lambda}} (d X_{\Lambda}) =
\exp(- H^{\Phi^t}_{\Lambda \setminus i,\emptyset}(X_{\Lambda \setminus i})) \bigotimes_{j \in \Lambda \setminus i} \rho^{\odot, x_j}(d X_j) \otimes \rho^{\odot, x_i}(d X_i)
\end{equation*}
So far we have established a nice representation for $Q^{x_{\Lambda}}_{\Lambda,y}$ with a new interaction $\Phi^{\varphi, t}$. The terms in the expression $\Phi^{\varphi, t}_A$ are of finite range and the derivatives of $\varphi$ are bounded, so $\Phi^{\varphi, t}$ is of finite range too.
Let us repeat here that we wrote the measure $Q_{\Lambda,y}^{x_{\Lambda}}$ (resp. $Q_{\Lambda, i, y}^{x_{\Lambda}} $) as a measure w.r.t. a product of independent Brownian motions on the circle. All dependencies are now shifted to the term $\exp(- H^{\Phi^t}_{\Lambda,\emptyset}(X_{\Lambda}))$ which we will control using a cluster expansion. Our next goal will be to show that those dependencies are small if $t$ is small. Due to its form it is clear that there exists a constant $C > 0$ such that for any $X$ and $A \subset \mathbb{Z}^d$
\begin{equation}
|\Phi^{\varphi, t}_A(X)| \leq C \biggl ( t + \sup_{j \in A} |X_j(t) - X_j(0)| \biggr ). \label{Phibound}
\end{equation}
Let us turn back to the Radon-Nikod\'ym derivative given by lemma $\ref{lemmaRN}$. We will now perform a cluster expansion of the nominator
\begin{equation}
\mathbb{E}_{Q_{\Lambda}^{x_{\Lambda}}} \biggl (\exp(f_{\Lambda,y}(X_{\Lambda})(t) - f_{\Lambda,y}(x_{\Lambda})) \biggr ) \label{cluexp}
\end{equation}
The one for the decoupled measure works analogously. Thanks to $\eqref{Girsanov}$ we can express the above expression $\eqref{cluexp}$ as
\begin{equation*}
\mathbb{E}_{\underset{j \in \Lambda} \bigotimes \rho^{\odot, x_j}} \biggl ( \exp \biggl ( - \sum_{A \subset \Lambda} \Psi_{A}^{y, \Lambda,t}(X_{\Lambda}) \biggr) \biggr )
\end{equation*}
where $\Psi_{A}^{y, \Lambda,t}$ is the interaction potential on $C([0,2\pi))^{\Lambda}$ given by
\begin{eqnarray*}
\Psi_{A}^{y, \Lambda, t}(X_{\Lambda}) & = & \Phi_A^{\varphi, t}(X) - \varphi_A(X(t)) + \varphi_A(X(0)) + \\
& & \sum_{B \cup \Lambda = A}\biggl ( \overset{\sim}\varphi_A(X_{\Lambda}(t)y_{\Lambda^c}) - \overset{\sim}\varphi_A(X_{\Lambda^c}(0)y_{\Lambda^c}) \biggr ).
 \end{eqnarray*}
Note that $\Psi_{A}^{y, \Lambda, t}$ satisfies the same bound as in $\eqref{Phibound}$.
Furthermore we remark that Girsanov gave us the representation of the expectation w.r.t. a product measure. \\

Now we want to turn to the space-time cluster expansion and need some more notation. Let $N \in \mathbb{N}$ be such that for $\sharp A > N$ one has $\Psi_{A}^{y, \Lambda,t} \equiv 0$ and let $\Gamma \subset \mathbb{Z}^d$ be finite s.t. for any $A$ with $\Psi_{A}^{y, \Lambda, t} \neq \emptyset$ one has $A \subset \underset{j \in A} \cap (\Gamma +j) $. We define a subclass of finite subsets in $\mathbb{Z}^d$, which play the role of polymers
\begin{equation*}
\mathcal{D}:= \lbrace A \subset \mathbb{Z}^d: 1 \leq \sharp A \leq N \text{ and } A \subset \underset{j \in A} \cap (\Gamma +j) \rbrace
\end{equation*}
Let $A_1,...A_n \in \mathcal{D}$. We call $\gamma = \lbrace A_1,..., A_n \rbrace$,
a collection of $n$ sets, a \textbf{cluster}. A cluster is called connected if for any $A_i, A_j \in \gamma$ there is a sequence $i=i_1,..,i_m=j$ s.t. $A_{i_1} \cap A_{i_2} \neq \emptyset ...A_{i_{m-1}} \cap A_{i_m} \neq \emptyset$. The set $\mathcal{A}$ will denote the set of all connected clusters (resp. $\mathcal{A}_{\Lambda}$ the set of connected clusters whose support is contained in $\Lambda$). Two clusters are called compatible if their supports are disjoint. $\mathcal{L}_{\Lambda}$ is the set of all compatible clusters. Finally we define the set $\mathcal{M}_{\Lambda}$ of collections of clusters such that their union is connected too.
We expand
\begin{eqnarray*}
\mathbb{E}_{\underset{j \in \Lambda} \bigotimes \rho ^{\odot, x_j}}\biggl ( \exp(- \underset{A \subset \Lambda} \sum \Psi_{A}^{ y, \Lambda,t}(X_{\Lambda})) \biggr ) & = & \\
\mathbb{E}_{\underset{j \in \Lambda} \bigotimes \rho ^{\odot, x_j}} \biggl (1 + \underset{n \geq 1} \sum \underset{\lbrace \gamma_1,..,\gamma_n \rbrace \in \mathcal{L}_{\Lambda}} \sum \prod_{i =1}^n \mathcal{K}^{y,\Lambda,t}(\gamma_i)(X) \biggr )
\end{eqnarray*}
where
\begin{equation*}
\mathcal{K}^{y, \Lambda,t}(\gamma_i)(X) = \underset{A \in \gamma_i} \prod \biggl (\exp(-\Psi_{A}^{y,\Lambda,t}(X_{\Lambda} )) - 1 \biggr) .
\end{equation*}
The clusters in $\mathcal{L}_{\Lambda}$ have disjoint supports, thus we can write the expectation of the product as a product of the expectations
\begin{equation*}
1 + \underset{n \geq 1} \sum \underset{\lbrace \gamma_1,..,\gamma_n \rbrace \in \mathcal{L}_{\Lambda}} \sum \prod_{i=1}^n \mathbb{E}_{\underset{j \in \Z^d} \bigotimes \rho ^{\odot, x_j}} ( \mathcal{K}^{y, \Lambda,t}(\gamma_i)(X ) ).
\end{equation*}
We will abbreviate $K_{x}^{y, \Lambda,t}(\gamma_i):= \mathbb{E}_{\underset{j \in \Z^d} \bigotimes \rho ^{\odot, x_j}} ( \mathcal{K}^{y, \Lambda,t}(\gamma_i)(X ))$, which plays the role of a ``weight" on the cluster. First we prove some estimates on these weights $K_{x}^{y, \Lambda,t}$.

\begin{Lemma}\label{Lemma_bigbound}
There exists a function $\lambda(t) > 0$ which tends to 0 as $t$ goes to 0, such that for any $y,x \in [0,2\pi)^{\Z^d}$, $\Lambda \subset \mathbb{Z}^d$ finite and connected clusters $\gamma \in \mathcal{A}$ we have the following estimate
\begin{equation*}
\mid K_{x}^{y, \Lambda, t}(\gamma) \mid \leq \lambda(t)^{\sharp \gamma}.
\end{equation*}
\end{Lemma}
\textbf{Proof:}\\
\begin{equation*}
\mid K_{x}^{y, \Lambda ,t}(\gamma) \mid = \biggl | \mathbb{E}_{\underset{j \in \Z^d} \bigotimes \rho ^{\odot, x_j}} \biggl ( \underset{A \in \gamma} \prod \exp (- \Psi_{A}^{y,\Lambda,t}(X_{\Lambda})) -1 \biggr ) \biggr |
\end{equation*}
Now we want to exchange the product and integration. We use lemma 5.2. from \cite{MinVerZag00} to write
\begin{equation*}
\biggl | \mathbb{E}_{\underset{j \in \Z^d} \bigotimes \rho ^{\odot, x_j}} \biggl ( \underset{A \in \gamma} \prod \exp (- \Psi_{A}^{y, \Lambda,t}(X_{\Lambda})) -1 \biggr ) \biggr | \leq \underset{A \in \gamma} \prod \mathbb{E}_{\underset{j \in \Z^d} \bigotimes \rho ^{\odot, x_j}} \biggl ( \biggl | \exp (- \Psi_{A}^{y, \Lambda, t}(X_{\Lambda})) -1 \biggr |^p \biggr )^{\frac{1}{p}}
\end{equation*}
for $p$ bigger than $N$ and an even number. We use $\eqref{Phibound}$ to estimate
\begin{eqnarray*}
\mathbb{E}_{\underset{j \in \Z^d} \bigotimes \rho ^{\odot, x_j}} \biggl( \biggl | \exp (- \Psi_{A}^{y, \Lambda , t}(X_{\Lambda})) -1 \biggr |^p \biggr )^{\frac{1}{p}} & \leq & \\
\mathbb{E}_{\underset{j \in \Z^d} \bigotimes \rho ^{\odot, x_j}} \biggl ( \biggl | \exp \biggl [ C \cdot t + \sup_{j \in A} \mid X_j(t) - X_j(0) \mid \biggl ] -1 \biggr |^p \biggr)^{\frac{1}{p}}
\end{eqnarray*}
We are seeking to get some bound on the $p$-th moment of an exponential function of a Brownian motion on the circle w.r.t. the N-dimensional Wiener measure on the circle. We recall that $N$ was chosen in such a way that for $\sharp A > N$ it follows $\Psi_A^{y, \Lambda, t}\equiv 0$. This moment exists and is equal to some positive constant $\lambda$ depending on the time $t$. Thus
\begin{equation*}
\underset{A \in \gamma} \prod \mathbb{E}_{\underset{j \in \Z^d} \bigotimes \rho ^{\odot, x_j}} \biggl( \biggl | \exp (- \Psi_{A}^{y, \Lambda,t}(X_{\Lambda})) -1 \biggl |^p \biggl)^{\frac{1}{p}} \leq \lambda(t)^{\sharp \gamma}.
\end{equation*}
From the bound $\eqref{Phibound}$ for the interaction $\Psi$ we deduce that $\lambda(t)$ tends to 0 as $t$ goes to 0.
\begin{flushright}
$\square$
\end{flushright}
Now we can deduce from the previous lemma that for $t$ small enough and incompatible clusters $\gamma \in \mathcal{A}$
\begin{equation}
\sup_{x,y \in [0, 2 \pi]^{\Z^d}} \text{ }\sup_{\Lambda \subset \Z^d} \sum_{\substack{\gamma^{\prime} \in \mathcal{A} : \\ supp (\gamma) \cap supp (\gamma^{\prime}) \neq \emptyset}} \mid K_{x}^{y, \Lambda, t}(\gamma^{\prime}) \mid \exp (\sharp \gamma^{\prime} ) \leq \sharp \gamma . \label{critRoe}
\end{equation}
In the following we use the Kotecky-Preiss condition from \cite{KotPre86}. In that paper the authors prove that if there are two positive functions $a$ and $d$ on the set of polymers such that one has for the polymer weights $\Phi$ and incompatible clusters $C$ the following inequality
\begin{equation}
\sum_{C, C^{\prime} \text{ incomp. }} \exp (a(C^{\prime}) + d(C^{\prime})) \mid \Phi(C^{\prime}) \mid \leq a(C) \label{criterium}
\end{equation}
then the log of the partition function admits a unique cluster expansion. The criterion $\eqref{criterium}$ is satisfied by $\eqref{critRoe}$ for $d=0$ and $a=|\cdot|$. Let us mention that the convergence criteria could be improved in \cite{FerPro07}. \\
Thus we can apply the cluster expansion following Kotecky and Preiss and get a unique expansion of the logarithm of the partition function.

\begin{eqnarray*}
\ln \biggl ( \mathbb{E}_{Q_{\Lambda}^{x_{\Lambda}}}(\exp( - \overset{\sim}H^{\overset{\sim}\varphi}_{\Lambda, \Lambda^c}(X_{\Lambda}(t),y) + \overset{\sim}H^{\overset{\sim}\varphi}_{\Lambda, \Lambda^c}(x,y)
)) \biggr ) & = & \\
\ln \biggl ( 1 + \underset{n \geq 1} \sum \underset{\lbrace \gamma_1,..\gamma_n \rbrace \in \mathcal{L}_{\Lambda}} \sum \prod_{i=1}^n K_{x}^{y,\Lambda, t}(\gamma_i) \biggr ) & = & \\
\underset{n \geq 1}\sum \underset{\lbrace \gamma_1,..\gamma_n \rbrace \in \mathcal{M}_{\Lambda}} \sum a(\gamma_1,..,\gamma_n)K_{x,t}^{y,\Lambda}(\gamma_1)..K_{x}^{y,\Lambda,t}(\gamma_n)
\end{eqnarray*}
with $a(\gamma_1,..,\gamma_n) \in \mathbb{R}$ coming from the Taylor expansion of the logarithm and where $\mathcal{M}_{\Lambda}$ was the set of connected clusters whose supports are connected too.
For the i-decoupled measure we have the same expression except we sum over $\mathcal{M}_{\Lambda \setminus i}$, the set of all clusters in $\Lambda \setminus i$ whose support is connected too. Therefore the log of the ratio $\frac{d \nu^t_{\Lambda, y}}{d \nu^{t,i}_{\Lambda, y}}(x_{\Lambda})$ provides
\begin{eqnarray}
\underset{n \geq 1}\sum \underset{\lbrace \gamma_1,..\gamma_n \rbrace \in \mathcal{M}_{\Lambda}} \sum a(\gamma_1,..,\gamma_n)K_{x}^{y,\Lambda,t}(\gamma_1)..K_{x}^{y,\Lambda,t}(\gamma_n) & - & \nonumber \\
\underset{n \geq 1}\sum \underset{\lbrace \gamma_1,..\gamma_n \rbrace \in \mathcal{M}_{\Lambda \setminus i}} \sum a(\gamma_1,..,\gamma_n)K_{x}^{y,\Lambda,t}(\gamma_1)..K_{x}^{y,\Lambda,t}(\gamma_n) & = & \nonumber \\
\underset{n \geq 1}\sum \sum_{\substack{\lbrace \gamma_1,..\gamma_n \rbrace \in \mathcal{M}_{\Lambda}: \\ i \in supp(\cup _j \gamma_j)}} a(\gamma_1,..,\gamma_n)K_{x}^{y,\Lambda,t}(\gamma_1)..K_{x}^{y,\Lambda,t}(\gamma_n) \label{series}
\end{eqnarray}
for $t$ small. The bound in $\eqref{critRoe}$ is uniform in $x,y$ and $\Lambda$, thus again using \cite{KotPre86} we can conclude that the former series $\eqref{series}$ converges uniformly in $x,y$ and $\Lambda$.
Take an arbitrary connected cluster $\gamma$, then
\begin{eqnarray*}
\mathbb{E}_{\underset{j \in \Lambda} \bigotimes \rho ^{\odot, x_j}}\biggl (\underset{A \in \gamma} \prod
(\exp(- \Psi_{A}^{y,\Lambda,t}(X_{\Lambda} ))-1) \biggr ) & \underset{\Lambda \longrightarrow \mathbb{Z}^d}
\longrightarrow & \\
\mathbb{E}_{\underset{j \in \mathbb{Z}^d} \bigotimes \rho ^{\odot, x_j}}\biggl (\underset{A \in \gamma} \prod (\exp(- \Psi^t_{A} (X))-1) \biggr)
\end{eqnarray*}
and it follows using the Lebesgue dominated convergence theorem that $\frac{d \nu^t_{\Lambda,y}}{d\nu^{t,i}_{\Lambda,y}}(x_{\Lambda})$ converges uniformly in $x,y$ towards $\exp(-\overset{\sim}H^{\overset{\sim} \varphi}_i(x)) \exp(G^t_i(x))$ with
\begin{equation*}
G_i^t(x) = \underset{n \geq 1}\sum \underset{\substack{\lbrace \gamma_1,..\gamma_n \rbrace \in \mathcal{M}_{\mathbb{Z}^d}: \\ i \in supp(\cup _j \gamma_j)}} \sum a(\gamma_1,..,\gamma_n)K^t_{x}(\gamma_1)..K^t_{x}(\gamma_n).
\end{equation*}
Because of the weak convergence of $\nu^t_{\Lambda, y}$ towards $\nu^t$, as well as for the decoupled measures, one has for each $i$ that $\nu^t(dx) = \exp(- \overset{\sim}H^{\overset{\sim} \varphi}_i(x) + G^t_i(x))\nu^{t,i}(dx)$.
For $g$ a local bounded function from $[0,2\pi)^{\Delta}$ to $[0,2\pi)$

\begin{eqnarray*}
\int_{[0,2\pi)^{\mathbb{Z}^d}} g(x_{\Delta}) \nu^t(dx) & = & \lim_{\Lambda \rightarrow \mathbb{Z}^d} \int_{[0,2\pi)^{\Lambda}} g(x_{\Delta}) \nu^t_{\Lambda}(dx_{\Lambda}) \\ & = & \int_{[0,2\pi)^{\mathbb{Z}^d}} g(x_{\Delta}) \exp (- \overset{\sim}H^{\overset{\sim}\varphi }_i (x) + G^t_i(x)) \nu^{t, i} (dx)
\end{eqnarray*}
thus it follows that the measures $\nu^t(dx)$ and $ \exp \biggl (- \overset{\sim} H^{\overset{\sim}\varphi }_i(x) + G^t_i(x) \biggr) \nu^{t,i}(dx)$ coincide for each $i$ and that the RN-derivative is continuous.
\medskip

The product measure on $[0, 2\pi)^{\Lambda \setminus i}\times [0,2\pi)$, $\nu^{t, i}_{\Lambda, y}$, is the measure of the decoupled dynamics with decoupled initial condition. This projection on the $i$-th coordinate is the Haar measure $\frac{1}{2 \pi} dx$. Therefore $\nu^{t, i}$ is a product measure on $[0,2\pi)^{\mathbb{Z}^d \setminus i} \times [0,2\pi)$ with also the Haar measure as projection on the $i$-th coordinate.

\medskip

Altogether we get for $t$ small that $\nu^t$ is a Gibbs measure associated to the Haar measure as reference measure with interaction $\varphi^t$ given by

\begin{eqnarray*}
\varphi^t_A(x) & = & \overset{\sim}\varphi_A(x)-\underset{n \geq 1}\sum \underset{\substack{ \lbrace \gamma_1,..\gamma_n \rbrace \in \mathcal{M}_{A}: \\ A= supp(\cup _j \gamma_j)}} \sum a(\gamma_1,..,\gamma_n)K^t_{x}(\gamma_1)..K^t_{x} (\gamma_n).
\end{eqnarray*}
We conclude that for $t$ small the time-evolved interaction is a small perturbation of the initial interaction. There exists a time-evolved interaction $\varphi^t$ which is given by the above equation and depends on $\overset{\sim} \varphi$ and $\varphi$ and is absolute summable, since $\overset{\sim} \varphi$ and $\varphi$ are absolutely summable. For the case that the initial measure $\nu$ is an extremal Gibbs measure we have proven that $\nu^t$ (associated to $\varphi^t$) is Gibbs too, i.e. $\nu^t \in \mathcal{G}(\varphi^t, \frac{1}{2 \pi}dx)$. It is even extremal in the set $\mathcal{G}(\varphi^t, \frac{1}{2 \pi}dx)$.

\medskip

An arbitrary Gibbs measure is a mixture of extremal Gibbs measures so if we take any average over extremal Gibbs measures we get the result for a general $\nu$.
\begin{flushright}
$\square$
\end{flushright}
\subsection{High temperatures at arbitrary times}

Now we look at arbitrary times $t$. We consider the infinite-dimensional gradient system where both the initial and the dynamical interactions are small. Let us recall the definition of the system
\begin{eqnarray}
\begin{cases}
& d X_i (t) = - \frac{1}{2} U^{\prime}(X_i(t))dt - \frac{\beta}{2}\nabla_i H^{\varphi}_i(X(t)) dt + d B_i^{\odot}(t), \text{ } i \in \mathbb{Z}^d, t > 0 \label{dynamics_largetime} \\
& X(0) \simeq \nu , \text{ } t=0
\end{cases}
\end{eqnarray}
and for $\beta = 0$
\begin{eqnarray}
\begin{cases}
& d X_i (t) = - \frac{1}{2} U^{\prime}(X_i(t))dt + d B_i^{\odot}(t) , \text{ } i \in \mathbb{Z}^d, t > 0 \label{dynamics_largetime22} \\
& X(0) \simeq \nu , \text{ } t=0.
\end{cases}
\end{eqnarray}
We note that the only difference with the previous section is that here the single-site term $U$ is not included in $H^{\varphi}$ but considered separately.
The initial interaction $\overset{\sim} \varphi$ will always be a "high-temperature interaction", so it satisfies condition $\eqref{Dobrushin}$. Therefore we will always start in a unique Gibbs measure $\nu$. Furthermore we distinguish, like in the previous part, between the case where
the dynamical interaction is infinite-temperature, like system $\eqref{dynamics_largetime22}$, which means that the evolution follows independent Brownian motions with drift moving on circles, and the system defined in $\eqref{dynamics_largetime}$ including both high- and infinite-temperature dynamics. We assume that $U$ is at least $C^2([0,2\pi))$.
\begin{Theorem}
Let $U$ be a $C^2$-function, let $\overset{\sim} \varphi$ be a high- or infinite-temperature interaction and let the finite-dimensional dynamical interaction $\varphi_{\Lambda}$ be $C^2$, for all $\Lambda$. Let $\beta$ be the inverse temperature. Furthermore let $Q^{\nu}_{\beta}$ be the law of the solution of $\eqref{dynamics_largetime}$ on $\overset{\sim} \Omega$ with the unique $\nu \in \mathcal{G}(\overset{\sim} \varphi, m)$ and $m$ given by $m(dx) = \frac{1}{Z}e^{-U(x)} \frac{1}{2 \pi} dx$. Then there exists an inverse temperature $\beta_0:=\beta_0(\overset{\sim}\varphi,\varphi) > 0$ such that for any $\beta \leq \beta_0$ and any $t$ there exists an interaction $\varphi^t$ which is absolutely summable, and which has as a Gibbs measure $\nu^t_{\beta} = Q^{\nu}_{\beta} \circ X(t)^{-1} \in \mathcal{G}_{\beta}( \varphi^t, m)$.
\end{Theorem}
\textbf{Proof:}\\
Here again the proof follows essentially the one of $\cite{DerRoe05}$. Note that in our case the ultracontractivity condition on $U$ is automatically satisfied, and the same holds for the balance condition on $U$ and the dynamical potential $\varphi$, namely
\begin{equation*}
\sup_{\Lambda \subset \mathbb{Z}^d}\sup_{i \in \Lambda} \sup_{x \in [0,2\pi)^{\Lambda}} |U^{\prime}(x_i) \cdot \nabla_i \varphi_{\Lambda}(x)| < \infty.
\end{equation*}
Ultracontractivity assures an exponential fast convergence of the system to equilibrium.
Since now we want to look at arbitrarily large times and are not restricted to a perturbative result around the initial measure, we need to use different techniques, similarly to the authors of \cite{DerRoe05}.
A sketch of the proof follows. Let us fix some notation first.

As defined before, we recall that $Q^{\nu}_{\beta}$ is the law of the solution of $\eqref{dynamics_largetime}$, depending on $\beta$ and the initial distribution $\nu$. Furthermore $Q^{x_{\Lambda}}_{\beta, \Lambda}$ will denote the law of the solution of the corresponding finite-dimensional problem on $\Lambda$ with initial condition $x_{\Lambda}$.
Analogously, let $Q^{\nu}_{0}$ denote the law of the solution of $\eqref{dynamics_largetime22}$ (the infinite-temperature dynamics) and $Q^{x_{\Lambda}}_{0, \Lambda}$ the finite-dimensional version.
\medskip

In the first step we compute the density of $Q^{x_{\Lambda}}_{\beta, \Lambda}$ with respect to $Q^{x_{\Lambda}}_{0, \Lambda}$, so we compare the difference evolving with high-resp. infinite-temperature dynamics after some time, again first for a finite box $\Lambda$. Again using cluster expansion techniques, with $\beta$ as the small parameter, we will be able to show that the Radon-Nikod\'ym derivative of these measures will turn out to be small. So it does not make much difference if the system evolves with high- or infinite-temperature dynamics. Next we will study the Gibbsian character of the joint distribution $\textbf{Q}_{\beta}^{\nu} = Q_{\beta}^{\nu} \circ (X(0),X(t))^{-1}$ on the space $C(S)$ instead of $\nu^t$ directly. This measure will turn out to be Gibbs. It is associated to some interaction $\textbf{H}^t$ which will depend on the initial Hamiltonian, the terms from the cluster expansion and a two-body potential induced by $p_t^{\odot, U}$ (the transition probability of a Brownian motion on the circle with drift $U^{\prime}$ which depends only on the single sites $i$). By integrating out the first layer we will be able to show the existence of of a regular density $f^t_{\Lambda,\beta}$ for which the conditional probabilities
\begin{equation*}
\nu_{\beta}^t(d z_{\Lambda} | y_{\Lambda^c}) = f_{\Lambda,\beta}^t(z_{\Lambda} y_{\Lambda^c}) m^{\otimes \Lambda}(d z_{\Lambda})
\end{equation*}
exist. Using Kozlov's representation theorem $\cite{Koz74}$ we will identify this measure to be Gibbs.

\medskip

Let us consider the time interval $[0,t]$ and let $\Lambda \subset \Z^d$. Similarly as in the previous section using Girsanov's theorem and It\^o's formula we get that $Q^{x_{\Lambda}}_{\beta, \Lambda}$ is absolutely continuous w.r.t. $Q^{x_{\Lambda}}_{0, \Lambda}$ with density
\begin{equation*}
F_{\beta, \Lambda}^t(X_{\Lambda}) =
\exp \biggl ( -\frac{1}{2} \beta H^{\varphi}_{\Lambda, \emptyset}(X_{\Lambda}(t)) + \frac{1}{2}\beta H^{\varphi}_{\Lambda, \emptyset}(X_{\Lambda}(0)) + \int_0^t \sum_{A \subset \Lambda} g_{\beta, A}^{U, \varphi}(X(s)) ds \biggr )
\end{equation*}
where the function $g_{\beta, A}^{U, \varphi}$ is $\mathcal{F}_A$-measurable and depends on the inverse temperature $\beta$, the drift term $U$ and dynamical interaction $\varphi$. It is defined as follows
\begin{eqnarray}
g_{\beta, A}^{U, \varphi}(x) & = & \frac{1}{4} \beta \sum_{i \in A} \biggl ( \frac{d^2}{d x_i^2} \varphi_A(x_A) + U^{\prime}(x_i) \frac{d}{d x_i} \varphi_A(x_A) \biggr) \nonumber \\
& & - \frac{1}{8} \beta^2 \sum_{\substack{B \cup C = A \\ B \cap C \neq \emptyset}} \sum_{i \in B \cap C} \frac{d}{d x_i} \varphi_B(x_B)\frac{d}{d x_i} \varphi_C(x_C).\label{defG}
\end{eqnarray}
So Girsanov's theorem and It\^o's formula yield the representation
\begin{equation*}
Q^{x_{\Lambda}}_{\beta, \Lambda} (d X_{\Lambda}) = F^t_{\beta, \Lambda}(X_{\Lambda}) Q^{x_{\Lambda}}_{0, \Lambda} (d X_{\Lambda}),
\end{equation*}
similarly as in the previous section.
We set the initial and final values to $X_{\Lambda}(0)=x_{\Lambda}$ and $X_{\Lambda}(t)=y_{\Lambda}$ and deduce from the previous considerations that
\begin{eqnarray}
& & \frac{d Q^{x_{\Lambda}}_{\beta, \Lambda} \circ X(t)^{-1}}{d Q^{x_{\Lambda}}_{0, \Lambda}\circ X(t)^{-1}}(y_{\Lambda}) = \label{RNdens} \\
& & \mathbb{E}_{Q^{x_{\Lambda}}_{0, \Lambda}} \biggl ( \exp \biggl(
-\frac{1}{2} \beta H^{\varphi}_{\Lambda, \emptyset}(X_{\Lambda}(t)) + \frac{1}{2}\beta H^{\varphi}_{\Lambda, \emptyset}(X_{\Lambda}(0)) + \int_0^t \sum_{A \subset \Lambda} g_{\beta, A}^{U, \varphi}(X(s)) ds \biggr) \biggr ) = \nonumber \\
& & e^{-\frac{1}{2} \beta [H^{\varphi}_{\Lambda, \emptyset}(y_{\Lambda}) - H^{\varphi}_{\Lambda, \emptyset}(x_{\Lambda})]}\mathbb{E}_{Q^{x, y}_{0, \Lambda}} \biggl ( \exp \biggl( \int_0^t \sum_{A \subset \Lambda} g_{\beta, A}^{U, \varphi}(X(s)) ds \biggr) \biggr ) \label{logCL}
\end{eqnarray}
We abbreviate the second factor by
\begin{equation}
\textbf{f}_{\Lambda, \beta}^t(x,y) := \mathbb{E}_{Q^{x_{\Lambda}}_{0, \Lambda}} \biggl ( \exp \biggl( \int_0^t \sum_{A \subset \Lambda} g_{\beta, A}^{U, \varphi}(X(s)) ds \biggr) \biggr ). \label{Rdens}
\end{equation}
To control the above term and therefore $\eqref{logCL}$ we have to perform a cluster expansion of $\textbf{f}_{\Lambda, \beta}^t(x,y)$ for $\beta$ small, i.e. we are in the case of high-temperature dynamics. It will be essentially the same as that of \cite{DerRoe05}. We want to work at the space-time level so a cluster will be a collection of subsets of $\Z^d \times \mathbb{N}$. The definition of compatibility and connectivity of clusters will be identical. For a number $M \in \mathbb{N}$ which we will fix later, we look at clusters whose supports are included in $\Lambda \times \lbrace 0,...,M \rbrace$. In fact $M=M_{\beta}^t$ will depend on time $t$ at which we are looking and on $\beta$. The time interval $[0,t]$ will also be discretized and divided into $M$ subintervals $[0,t] = \bigcup_{i=0}^{M-1}[j\frac{t}{M}, (j+1)\frac{t}{M}] = \bigcup_{i=0}^{M-1} I_i$. Let $p_t^{\odot, U}(x_i,\cdot)$ be the kernel of the free dynamics, i.e. independent Brownian motion with drift $U^{\prime}$ w.r.t. $m$. Note if $U \equiv 0$ then $p_t^{\odot, 0}(x_i,\cdot) = p_t^{\odot}(x_i,\cdot)$, where $p_t^{\odot}$ was defined as the kernel of Brownian motion moving on a circle and starting in $x_i$.

\medskip

We obtain the following expansion of $\textbf{f}_{\Lambda, \beta}^t$
\begin{equation*}
\textbf{f}_{\Lambda, \beta}^t(x,y) = 1 + \sum_{n=0}^{\infty} \underset{\lbrace \gamma_1,...,\gamma_n \rbrace \in \mathcal{L}_{\Lambda \times [0,M]}} \sum \textbf{K}^{x,y, t}_{M,\beta}(\gamma_1)(X)...\textbf{K}^{x,y, t}_{M,\beta}(\gamma_n)(X) 
\end{equation*}
where
\begin{equation*}
\textbf{K}^{x,y, t}_{M,\beta}(\gamma_i)(X) = \int ... \int \mathcal{K}^t_{M,\beta}(\gamma_i)(X)Q^{x,x^{(1)}}_{0, \Lambda, I_0}(dX)... Q^{x^{(M-1)}, y}_{0, \Lambda, I_{M-1}}(dX)m^{\otimes \Lambda}(dx^{(M-1)}_{\Lambda}) .
\end{equation*}
$Q^{x,x^{(1)}}_{0, \Lambda, I_0}$ denotes the law of the solution of the free ($\beta = 0$) dynamics in the time interval $I_0$ starting at time 0 in $x$ and being at time $t/M$ in $x^{(1)}$. $\mathcal{K}^t_{M,\beta}(\gamma_i)$ consists of products on terms which depend on the $p_t^{\odot, U}$ and $g_{\beta,A}^{U, \varphi}$ on subintervals of $[0,t]$. We want to use the Kotecky-Preiss condition, as in the section before, to obtain the expansion of the log of $\textbf{f}_{\Lambda, \beta}^t$. Therefore we need analogously that there exists a function $\lambda(\beta)$ tending to 0 for $\beta$ going to infinity such that
\begin{equation*}
\sup_{x,y,t} |\textbf{K}^{x,y, t}_{M,\beta}(\gamma_i)| \leq \lambda(\beta)^{\sharp \gamma_i},
\end{equation*}
which follows from
\begin{equation*}
|\mathcal{K}^t_{M, \beta}(\gamma) | \leq \lambda(\beta)^{\sharp \gamma_i}.
\end{equation*}
Let $T:=t/M$. To prove the last statement we define the maximal fluctuation of the kernel w.r.t. the stationary measure $p_t^{\odot, U}$ on $[T,\infty)$ around te equilibrium
\begin{equation*}
\epsilon(T) = \sup_{t \geq T} \sup_{a,b \in [0,2\pi)}|p_t^{\odot, U}(a,b) - 1|
\end{equation*}
and know, since $U$ is nicely defined and the spins are compact, that for $T$ going to infinity, $\epsilon(T)$ is going to 0. We choose $\beta_0$ such that for any $\beta < \beta_0$ there exists a constant $C$ with
\begin{equation*}
(1 + \epsilon(1/2\sqrt{\beta_0}))e^{C \sqrt{\beta_0}} - 1 \leq 1.
\end{equation*}
For $\beta \leq 1/t^2$, we set $M=1$ so $T=t$ and therefore the bound for the cluster weights will be $\lambda(\beta) = \biggl( e^{C_1 \sqrt{\beta}} - 1 \biggr )^{\frac{| \gamma |}{2 \sharp \Gamma}}$. For the other case when $\beta > 1/t^2$ we set $T: = t / [t\sqrt{\beta}] +1 $ and get for the bound $\lambda(\beta) = \biggl( (1+\epsilon(1/2\sqrt{\beta}) )e^{C_2 \sqrt{\beta}} - 1 \biggr )^{\frac{| \gamma |}{2 \sharp \Gamma}}$.
Now we can use the Kotecky-Preiss condition and get an expansion for the logarithm of $\textbf{f}^t_{\Lambda, \beta}$. Let $Tr(\gamma_1,..,\gamma_n) \subset \Z^d$ be the projection of the cluster $\gamma_1,...,\gamma_n$ on its support. Then we can write the log in terms of projections at time 0 and $t$ coming down from the space-time level, namely
\begin{equation*}
\ln \biggl ( \textbf{f}_{\Lambda, \beta}^t(x,y) \biggr ) = \underset{\Delta \subset \Lambda} \sum \sum_{n=0}^{\infty} \sum_{\substack{\lbrace \gamma_1,...,\gamma_n \rbrace \in \mathcal{M}_{\Lambda \times [0,M]} \\ Tr(\gamma_1,...,\gamma_n)=\Delta} } a(\gamma_1,...,\gamma_n) \textbf{K}_{M, \beta}^{x,y,t}(\gamma_1)...\textbf{K}_{M,\beta}^{x,y,t}(\gamma_n).
\end{equation*}
Next, we express the previous RN-derivative
\begin{equation*}
\frac{d Q^{x_{\Lambda}}_{\beta, \Lambda} \circ X(t)^{-1}}{d Q^{x_{\Lambda}}_{0, \Lambda}\circ X(t)^{-1}}(y_{\Lambda}) = e^{-\frac{1}{2} \beta [H^{\varphi}_{\Lambda, \emptyset}(y_{\Lambda}) - H^{\varphi}_{\Lambda, \emptyset}(x_{\Lambda})]} e^{\log (\textbf{f}_{\Lambda, \beta}^t(x,y))}
\end{equation*}
in terms of the cluster expansion of the logarithm of $\textbf{f}_{\Lambda, \beta}^t(x,y)$ and get a nice form
\begin{equation*}
\frac{d Q^{x_{\Lambda}}_{\beta, \Lambda} \circ X(t)^{-1}}{d Q^{x_{\Lambda}}_{0, \Lambda}\circ X(t)^{-1}}(y_{\Lambda}) = e^{- \sum_{\Delta \subset \Lambda} \textbf{$\Phi$}_{\Delta}^{\beta, t}(x,y)}
\end{equation*}
with
\begin{eqnarray*}
\textbf{$\Phi$}_{\Delta}^{\beta, t}(x,y) & = & \frac{\beta}{2}( \varphi_{\Delta}(y) - \varphi_{\Delta}(x)) - \\
& & \sum_{n=0}^{\infty} \sum_{\substack{\lbrace \gamma_1,...,\gamma_n \rbrace \in \mathcal{M}_{\Lambda \times [0,M]} \\ Tr(\gamma_1,...,\gamma_n)=\Delta} } a(\gamma_1,...,\gamma_n) \textbf{K}_{M, \beta}^{x,y,t}(\gamma_1)...\textbf{K}_{M, \beta}^{x,y,t}(\gamma_n).
\end{eqnarray*}
For $\beta$ small enough it is easy to see that $\textbf{$\Phi$}^{\beta, t}$ is a "high-temperature interaction", i.e. satisfies the Dobrushin condition $\eqref{Dobrushin}$.
\medskip

Next, we investigate the joint distribution on the bi-space, $\textbf{Q}^{\nu}_{\beta} = Q_{\beta}^{\nu} \circ (X(0),X(t))^{-1}$. The Hamiltonian of the measure will additionally depend on the terms $\textbf{$\Phi$}_{\Delta}^{\beta, t}(x,y)$. We can prove that $\textbf{Q}^{\nu}_{\beta}$ is Gibbs is unique on the bi-space.
\begin{Lemma}
The measure $\textbf{Q}_{\beta}^{\nu}$ is a Gibbs measure on the bi-space associated to the reference measure $m \times m$ on $[0,2\pi)\times[0,2\pi)$ and formal Hamiltonian function $\textbf{H}^t$ given by
\begin{equation}
\textbf{H}^t_{\bigtriangleup}(x,y) = \overset{\sim} H^{\overset{\sim} \varphi}_{\bigtriangleup}(x) - \sum_{i \in \bigtriangleup } \log (p_t^{\odot, U}(x_i, y_i)) + \sum_{\substack{A \subset \Z^d \\ A \cap \bigtriangleup \neq \emptyset}} \textbf{$\Phi$}_{A}^{\beta, t}(x,y) \label{biH}.
\end{equation}

\end{Lemma}
The main argument uses the fact that both $\overset{\sim} \varphi$ and $\textbf{$\Phi$}^{\beta, t}$ are high-temperature interactions for sufficiently low $\beta$, so that the set of Gibbs measures on $([0,2\pi) \times [0,2\pi))^{\Z^d}$ associated to $\textbf{H}^t$ and reference measure $m \times m$ contains at most one element. Taking the natural bijection between $([0,2\pi) \times [0,2\pi))^{\Z^d}$ and $S$ we obtain the desired result.
\medskip

Now, let $\Lambda \subset \Z^d$ be a finite box. We fix the boundary condition in the second layer $X_{\Lambda^c}(t) = y_{\Lambda^c}$ and note that this measure $\textbf{Q}^{\nu, y_{\Lambda^c}}_{\beta}:= \textbf{Q}^{\nu}_{\beta}(\cdot, \cdot \mid X_{\Lambda^c}(t) = y_{\Lambda^c})$ is Gibbs on $[0,2\pi)^{\Z^d \times \lbrace 0 \rbrace \cup \Lambda \times \lbrace 1 \rbrace}$. Desintegrating the w.r.t. the first layer we get that for almost all $y_{\Lambda^c}$ , $\nu_{\beta}^t(\cdot | y_{\Lambda^c})$ is the marginal of $\textbf{Q}^{\nu, y_{\Lambda^c}}_{\beta}$. It is easy to see that there exists a regular density $f_{\Lambda,\beta}^t$ such that
\begin{equation*}
\nu_{\beta}^t(dz_{\Lambda} \mid y_{\Lambda^c}) = f_{\Lambda,\beta}^t(z_{\Lambda}y_{\Lambda^c}) m^{\otimes \Lambda}(dz_{\Lambda})
\end{equation*}
for $y \in [0,2\pi)^{\Z^d}$ given by
\begin{equation*}
f_{\Lambda,\beta}^t(y) = \int_{[0,2\pi)^{\Z^d}}\frac{1}{\textbf{Z}_{\Lambda}(y_{\Lambda^c})}\prod_{i \in \Lambda} p_t^{\odot, U}(x_i,z_i) \exp \biggl( \sum_{\substack{A \in \Z^d \\ A \cap \Lambda \neq \emptyset}} \textbf{$\Phi$}_A^{\beta}(x, z_{\Lambda}y_{\Lambda^c}) \biggr ) \overline{Q}^{\nu, y_{\Lambda^c}}_{\beta}(dx),
\end{equation*}
for $\overline{Q}^{\nu, y_{\Lambda^c}}_{\beta}(dx)$ a probability measure on $[0,2\pi)^{\Z^d}$. Finally we can analogously demonstrate the assumptions for Kozlov's representation theorem, see \cite{Koz74}:
\begin{Lemma}
For any $\Lambda$, $f_{\Lambda,\beta}^t$ satisfies the following properties
\begin{enumerate}
\item $\exists C_1, C_2 > 0 : C_1 \leq \underset{y \in [0,2\pi)^{\mathbb{Z}^d}}\inf f_{\Lambda,\beta}^t(y) \leq \underset{y \in [0,2\pi)^{\mathbb{Z}^d}} \sup f_{\Lambda,\beta}^t(y) \leq C_2$ and
\item $\lim_{\Delta \rightarrow \Z^d} \sup_{y, u: y_{\Delta}= u_{\Delta}} \mid f_{\Lambda,\beta}^t(y) - f_{\Lambda,\beta}^t(u) \mid =0$
\end{enumerate}
\end{Lemma}
and thus the family of conditional probabilities is built on an absolutely summable interaction and is therefore Gibbs.

\medskip
\textbf{Remark:}\\
The authors of \cite{DerRoe05} prove in one of their corolloraries that the evolved measure is a unique Gibbs measure for the evolved interaction if either the time is large enough or for all times,when the system evolves with an infinite-temperature dynamics. The argument goes essentially the same with less complication in our case; it uses the following argument. We already noticed that for $\beta$ small enough and $t$ large enough the potential on $S$ associated to $\textbf{H}^t$, as defined in $\eqref{biH}$, is a high-temperature interaction because it satisfies the high-temperature Dobrushin condition $\eqref{Dobrushin}$. A fortiori the specifications for this Hamiltonian are global, see \cite{FerPfi97}, such that the DLR-equations also hold for unbounded subsets of the bi-space $S$. The uniqueness of the evolved measure follows then from this global property.

\begin{flushright}
$\square$
\end{flushright}

\section{Loss of the Gibbs property for the plane rotor in two dimensions}

In this section we investigate what happens with the Gibbs measure for long times if one starts from the classical plane rotor model at a sufficiently low temperature in $\mathbb{Z}^2$ and evolves with independent Brownian motions moving on circles. We show that we can find a "bad configuration" for the time-evolved measure, which implies that after some time Gibbsianness gets lost.
\newline

In the spirit of \cite{vEntFerHolRed02} and \cite{KueRed06} we consider the joint distribution of the spins at time 0 and at time $t$, $\textbf{Q}^{\nu_{\beta}}$ which we already encountered in the previous section. Then we condition on a particular configuration at time $t$, $y^{spec}$, and show that this conditioning will create a set of alternating magnetic fields $(h_i(t))_{i \in \Z^2}$, so the configuration of a spin at site $i$ at time $t$ induces a local magnetic field $h_i(t)$. At time $t=0$ these fields will provide two ground states and a phase transition (by breaking a discrete left-right symmetry) for the conditioned model for certain choices of configurations which make the model depend sensitively on variations of these configurations outside arbitrary large volumes. This "bad configuration", $y^{spec}$, implies now non-Gibbsianness for the measure $\nu^t_{\beta}$ because looking ``backwards" from time $t$ to time 0 we see that the discontinuity w.r.t. the boundary at time $t$ when one conditions on the ``bad value" comes from the existence of two distinct Gibbs measures for the conditioned system at time 0 (phase transition at time 0), compare definition \ref{badconfig}.
\newline

The process $X=(X_i(t))_{t \geq 0, i \in \Z^2}$ in this case is defined by the following SDE
\begin{eqnarray}
\begin{cases}
& d X_i(t) = d B_i^{\odot}(t) , i \in \mathbb{Z}^2, t > 0 \label{system2-1} \\
& X(0) \simeq \nu_{\beta}, t=0
\end{cases}
\end{eqnarray}
for $\nu_{\beta} \in \mathcal{G}_{\beta}(\overset{\sim} \varphi, \nu_0)$ and $\overset{\sim} \varphi$ of finite range and Lipschitz continuous and $\nu_0(dx) = \frac{1}{2\pi} dx$. We write the initial measure $\nu_{\beta}$ explicitly depending on the inverse temperature $\beta$, and we will consider the standard nearest-neighbour interaction, which has a unique translation-invariant pure Gibbs measure \cite{BriFonLan77}.
The main theorem we want to prove is the following.
\begin{Theorem}
Let $Q^{\nu_{\beta}}$ be the law of the solution $X$ of the planar rotor system $\eqref{system2-1}$ in $\Z^2$, $\nu_{\beta} \in \mathcal{G}_{\beta}(\overset{\sim}\varphi, \nu_0)$ and $\overset{\sim} \varphi$ given by $\overset{\sim} \varphi_{\beta, A}(x) = -\beta J \underset{i,j \in A: i \sim j} \sum \cos(x_i-x_j)$, $J$ some non-negative constant.
Then there is a time interval $(t_0,t_1)$ such that for $\beta$ large enough the measure $\nu_{\beta}^t=Q^{\nu_{\beta}}\circ X(t)^{-1}$ is not Gibbs, i.e. $\nu_{\beta}^t \notin \mathcal{G}_{\beta}(\varphi^t, \nu_0)$ or in other words: one cannot find any version of its conditional probabilities which is a continuous function of the boundary condition (failure of quasilocality).
\end{Theorem}
\textbf{Proof:}\\
The outline of the proof is as follows. Like in the previous section we look at the joint law of the process at time 0 and time $t$, $\textbf{Q}^{\nu_{\beta}}$, on
the bi-space $S$. We fix a particular configuration at time $t$, namely an alternating up-down configuration $y^{spec}$, and show that the conditioned Hamiltonian $\textbf{H}^t_{\beta}(x, y^{spec})$ has two ground states for $t$ large enough. 
By applying a percolation argument for low-energy clusters and discrete symmetry breaking from \cite{Geo81}, we will be able to prove that for sufficiently low temperatures $|\mathcal{G}_{\beta}(\textbf{H}^t_{\beta}(\cdot, y^{spec}), \nu_0)| \geq 2 $, a phase transition occurs for the conditioned model at time 0. This means we have found a "bad configuration" $y^{spec}$ for $\nu_{\beta}^t$ such that the measure will fail to be Gibbs. \newline
We will rewrite the formal joint Hamiltonian, originally given by
\begin{equation}
\textbf{H}^t_{\beta}(x, y) =
\overset{\sim} H^{\overset{\sim} \varphi}_{\beta}(x) + \sum_{i \in \Z^2} \log (p^{\odot}_t(x_i, y_i)) \label{origiHam}
\end{equation}
as
\begin{equation}
\textbf{H}^t_{\beta}(x, y) =
\overset{\sim} H^{\overset{\sim} \varphi}_{\beta }(x) + \sum_{i} \biggl ( 2e^{-t}\cos (x_i-y_i) + o(e^{-t}) \biggr ). \label{Hamiltonian}
\end{equation}
The kernel $p_t^{\odot}(x_i,\cdot)$ in $\eqref{origiHam}$ was already defined in the previous section as the transition kernel of a Brownian motion on the circle starting from $x_i$. It has the explicit form
\begin{equation*}
p^{\odot}_t(x_i,y_i) = \frac{1}{\sqrt{2\pi t}} \underset{n \in \Z} \sum \exp \biggl ( - \frac{(y - x - 2n\pi)^2}{2t} \biggr )
\end{equation*}
see for example \cite{Ros97}. Using the Poisson summation formula it equals
\begin{equation*}
p_t^{\odot}(x_i,y_i) = \frac{1}{2 \pi} \biggl ( 1 + 2\cdot \sum_{n=1}^{\infty} e^{-n^2 t} \cos(n(x_i-y_i)) \biggr).
\end{equation*}
We will use the latter for convenience. It allows us to neglect the
above single-site correction terms $o(e^{-t})$,in the Hamiltonian,
which are bounded uniformly in
$x_i$ and $y_i$ (by $ const \times e^{-4t}$ in fact), and have the obvious symmetries, without changing the qualitative behaviour in the rest of the analysis.
In the following we investigate the conditioned model.
We choose an alternating up-down configuration at time $t$, namely $y^{spec} =(\1_{\lbrace i \in 2\mathbb{N} + 1 \rbrace} \pi, i\in \mathbb{Z}^2)$, which will yield two ground states. For convenience we call a spin $x_i$ \textbf{up} if $x_i = 0$ and \textbf{down} if it has the value $\pi$. $x_i$ points to the right (resp. to the left ) if $x_i = \pi/2$ (resp. $3/2 \pi $). By abuse of notation we say $i=(i_1,i_2) \in \Z^2$ \textbf{even} if the sum of their components is even, i.e. $i_1 + i_2 \in 2\mathbb{N}$, and \textbf{odd} otherwise. For example, the spin at the origin takes the value $y^{spec}_{(0,0)} = y^{spec}_0 = 0$ and at position $(1,0)$ the value $y^{spec}_{(1,0)}= \pi$ and so forth. Then we get the following ground states.
\begin{propo}\label{ground state}
Let $\nu_{\beta}$ be a sufficiently low-temperature initial Gibbs measure with $\beta=\beta(t)$ of order at most $\mathcal{O}(h(t)^{-2})$. Let the Hamiltonian of the conditioned joint system at time $0$ and $t$, $\textbf{H}^t_{\beta}(x, y)$, be defined in $\eqref{Hamiltonian}$. Then we find the following 2 ground states for this Hamiltonian $x^{ri} := (\pi /2 + (-1)^{i}\e_t, i \in \mathbb{Z}^2)$ and $x^{le} := (3/2 \pi + (-1)^{i}\e_t, i \in \mathbb{Z}^2)$ where $\e_t > 0$ is of order $\mathcal{O}(h(t)/J\beta)$.
\end{propo}
\textbf{Proof:}\\
We plug $y^{spec}$ in $\eqref{Hamiltonian}$ and obtain for the formal Hamiltonian
\begin{equation*}
\textbf{H}^t_{\beta}(x, y^{spec}) =
\overset{\sim} H^{\overset{\sim} \varphi}_{\beta }(x) + \sum_{i} \biggl ( 2(-1)^{i}e^{-t}\cos (x_i) + o(e^{-t}) \biggr ).
\end{equation*}
The initial formal Hamiltonian was chosen as
\begin{equation*}
\overset{\sim} H^{\overset{\sim} \varphi}_{\beta }(x) = - \beta J \sum_{i \sim j}\cos(x_i - x_j) .
\end{equation*}
where $J > 0$ is some nearest neighbour coupling term, which makes the initial system ferromagnetic. Now, we want to determine the configurations minimizing the Hamiltonian $\textbf{H}^t_{\beta}(x, y^{spec})$. We will here neglect the correction term. A configuration $x$ then is called a \textbf{ground state} if for each $i \in \Z^2$ the pair $(x_i, x_{i+1})$ is a minimal point of the real function
\begin{equation*}
\Phi^t_{\beta}: (z,y) \rightarrow -\beta J \cos(z-y) + \frac{1}{4} h_{i}(t)(\cos(z) - \cos(y))
\end{equation*}
with $h_{i}(t):=2(-1)^{i}e^{-t}=(-1)^i h(t)$. We can safely forget about the correction term here, which is small with respect to the fields $h_i(t)$, and has the same (left-right) symmetry. Note that this interaction $\Phi^t_{\beta}$ is equivalent to the interaction of our Hamiltonian $\textbf{H}^t_{\beta}(x, y^{spec})$. In fact
it is a sum of two competing terms. The term coming from the initial Hamiltonian wants neighbouring spins to point in the same direction and the other term, which comes from the conditioning at time $t$, tries to direct the spins
in the specified up-down directions. They will find a compromise at pointing almost in the same direction with small corrections alternatingly "up" and "down" by the amount of $|\e_t|$.
Let us assume that $i$ is even. We take derivatives of $\Phi_{\beta}^t(z,y)$ w.r.t. $z$ and $y$,
\begin{eqnarray*}
\frac{\partial}{\partial z} \Phi^t_{\beta}(z,y) & = & \beta J \sin(z - y) - \frac{1}{4}h(t) \sin(z) \\
\frac{\partial}{\partial y} \Phi^t_{\beta}(z,y) & = & -\beta J \sin(z - y) + \frac{1}{4}h(t) \sin(y).
\end{eqnarray*}
Thus the point $(z,y)$ is stationary if
\begin{eqnarray*}
\beta J \sin(z-y) & = & \frac{1}{4}h(t) \sin(z) \text{ and } \\
\beta J \sin(z-y) & = & \frac{1}{4}h(t) \sin(y)
\end{eqnarray*}
i.e. if $\sin(z) = \sin(y)$ which is the case if either $z = y$ or $z + y = \pi$. If $z=y$ then it follows from the conditions above that the only possible points are $(0,0)$ and $(\pi,\pi)$. For the second case, if $z+y = \pi$, we need that
\begin{eqnarray*}
\frac{4 \beta J}{h(t)} \sin(2 z) & = & -\sin(z) \text{ and } \\
\frac{4 \beta J}{h(t)} \sin(2 y) & = & \sin(y)
\end{eqnarray*}
which is equivalent to
\begin{eqnarray*}
\sin(z) \biggl ( 1 + \frac{8 \beta J}{h(t)} \cos(z) \biggr ) & = & 0 ,\\
\sin(y) \biggl ( 1 - \frac{8 \beta J}{h(t)} \cos(y) \biggr ) & = & 0.
\end{eqnarray*}
So either $\sin(z)$ and $\sin(y)$ is equal to 0, which means that all the spins want to point in either 0 and $\pi$-direction ($(0,\pi)$ and $(0,\pi)$ ), or $\cos(z) = - \frac{h(t)}{8 \beta J}$ and $\cos(y) = \frac{h(t)}{8 \beta J}$. For the low-temperature regime, when $\beta$ is very large, in fact we will choose $\beta(t)=\beta$ at least of order $\mathcal{O}(h(t)^{-2})$ , and since $J > 0$, the term $\pm \frac{h(t)}{8 \beta J}$ approaches 0 from above, respectively from below. Therefore $(z, y)$ has to be equal to $(\frac{\pi}{2} + \epsilon_t,\frac{\pi}{2} - \epsilon_t)$ resp. to $(3\frac{\pi}{2} + \epsilon_t, 3\frac{\pi}{2} - \epsilon_t)$ for some $\epsilon_t > 0$ which depends on $h(t)$. It will be of order at most $\mathcal{O}(h(t) / \beta J)$ (or $\mathcal{O}(h(t)^3)$ since we chose $\beta$ at least $\mathcal{O}(h(t)^{-2})$. Note that the assumption "$i$ even" reflects itself in the sign of $\epsilon_t$. For $i$ odd the extremal points would be $(\frac{\pi}{2} - \epsilon_t, \frac{\pi}{2} + \epsilon_t)$ resp. $(3\frac{\pi}{2} - \epsilon_t,3 \frac{\pi}{2} + \epsilon_t)$. Now we want to determine which configurations are the proper minima of $\Phi^t_{\beta}$ and therefore of $\textbf{H}^t_{\beta}$. The second derivatives are
\begin{eqnarray*}
\frac{\partial^2}{\partial z^2} \Phi^t_{\beta}(z,y) & = & \beta J \cos(z - y) - \frac{1}{4}h(t) \cos(z) \\
\frac{\partial^2}{\partial y^2} \Phi^t_{\beta}(z,y) & = & \beta J \cos(z - y) + \frac{1}{4}h(t) \cos(y) \\
\frac{\partial^2}{\partial z \partial y} \Phi^t_{\beta}(z,y) & = & - \beta J \cos(z - y) = \frac{\partial^2}{\partial y \partial z} \Phi^t_{\beta}(z,y).
\end{eqnarray*}
The determinant of the Hessian matrix is equal to $-\frac{1}{16} h(t) \cos(z) \cos(y)$. It is strictly negative for the points $(0,0)$ and $(\pi,\pi)$ and
strictly positive for $(\pm \frac{\pi}{2} - \epsilon_t, \pm \frac{\pi}{2} + \epsilon_t)$ and also $(0,\pi)$ resp. $(\pi,0)$. Therefore $(0,0)$ and $(\pi,\pi)$ are saddle points. The latter two are maximal points since $\frac{\partial^2}{\partial z^2} \Phi^t_{\beta}(z,y)|_{(0,\pi)} = - \beta J - \frac{1}{4}h(t) < 0$ for large times ($\beta$ large enough). The same is true for the point $(\pi,0)$. Since
\begin{equation*}
\beta J \cos(2\epsilon_t) > \frac{1}{4}h(t)\sin(\epsilon_t) + o(e^{-t}),
\end{equation*}
which is true for $t$ large, it follows that
\begin{equation*}
\frac{\partial^2}{\partial z^2} \Phi^t_{\beta}(z,y)|_{(\pi/2 + \e_t,\pi/2 - \e_t)} > 0
\end{equation*}
and the same holds for the point $(3/2 \pi + \e_t, 3/2 \pi - \e_t)$. So these points are the only minima. For an illustration we show how the ground states look like.

\begin{minipage}[hbt]{5cm}
\centering
\includegraphics[width= 4 cm, height= 4 cm, angle= 270]{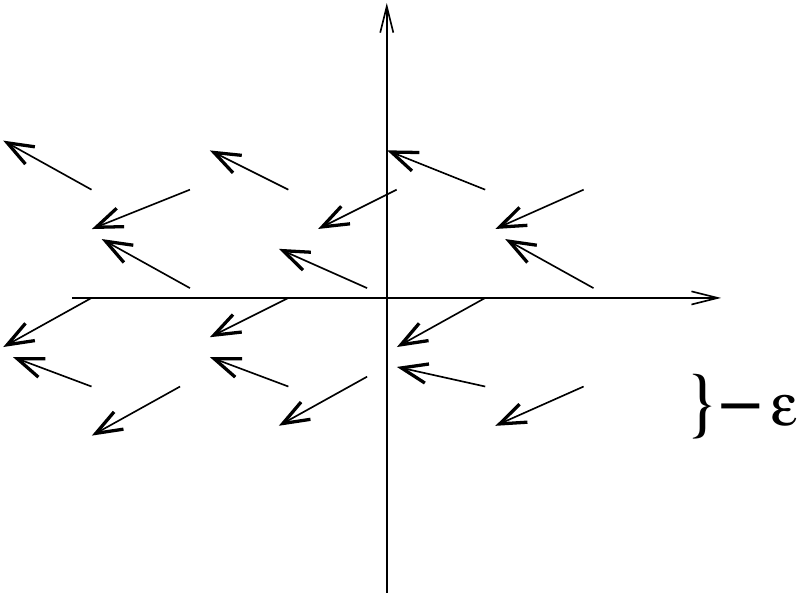}
$x^{le}$
\end{minipage}
\hfill
\begin{minipage}[hbt]{5cm}
\centering
\includegraphics[width= 4 cm, height= 4 cm, angle= 270]{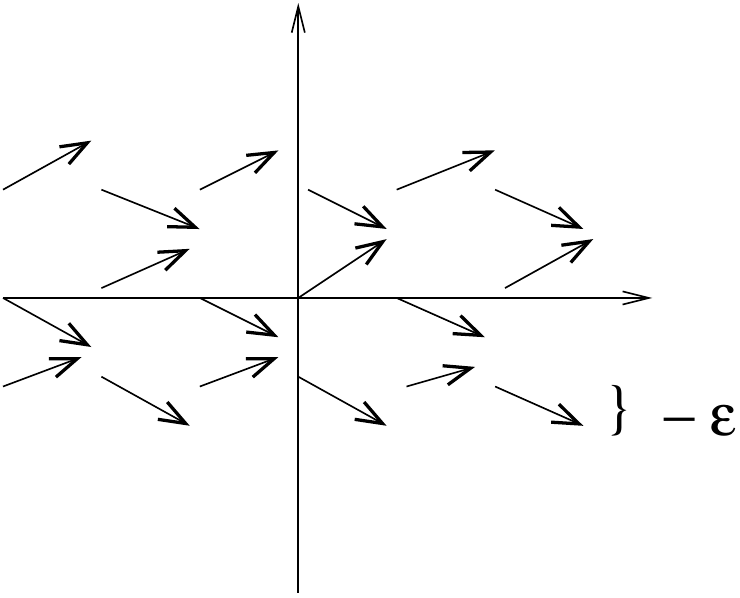}
$x^{ri}$
\end{minipage}

\begin{flushright}
$\square$
\end{flushright}
Now we are in a position to prove a phase transition at time $0$. We will essentially follow the arguments of \cite{Geo81}. By conditioning with the configuration $y^{spec}$ at time $t$ we created a discete left-right symmetry for the continuous model at time $0$. Now we want to prove via percolation methods that there is a time interval $(t_0,t_1)$ for which a spontanous symmetry breaking occurs.
We will look at Gibbs measures which are obtained by taking infinite-volume limits of finite-volume Gibbs measures with periodic boundary conditions. The picture of what happens is the following: if the interaction exhibits a ground state degeneracy then the "low-energy ocean" is bound to show a pattern corresponding to one of the 2 distinct ground states. This pattern can be seen at infinity and the 2 possible values are symmetry related. A fortiori there exists 2 disjoint symmetry-related tail events with positive probability, which can only occur if the relating symmetry is broken.

\medskip

One question Georgii answers in \cite{Geo81} is under which conditions there exists a unique "low-energy" infinite cluster and when it implies that the system has a phase transition. Two adjacent sites have a low-energy interaction if for a given $\delta > 0$, their energy is smaller than the ground state energy plus the correction term $\delta$. The existence of a low-energy bond percolation will be implied by a low-energy site percolation on the dual lattice for our nearest neighbour potential having the properties below.

\medskip

To use percolation for low-energy clusters on the dual lattice and Reflection Positivity arguments similar to \cite{Geo81} we need to check some assumptions on the interaction $\Phi^t_{\beta}$.

\medskip

Let $(\mathbb{Z}^2)^*$ indicate the dual lattice of the square lattice, i.e. $(\mathbb{Z}^2)^* = \mathbb{Z}^2 + (1/2, 1/2)$. Each point $a$ in $(\mathbb{Z}^2)^*$ is identified with the elementary cube consisting of 4 sites of $\mathbb{Z}^2$, i.e.
\begin{equation*}
a \equiv \lbrace i \in \mathbb{Z}^2 : i = a + (\pm 1/2, \pm 1/2) \rbrace.
\end{equation*}
Let $v=\lbrace (0,0), (0,1), (1,1), (1,0) \rbrace \equiv (1/2,1/2)$ be the elementary cube at the origin of $(\mathbb{Z}^2)^*$.
We will write the formal Hamiltonian $\textbf{H}^t_{\beta}$ as an Hamiltonian on the dual lattice, i.e. of the form $\textbf{H}^t_{\beta}(x) = \sum_{a \in (\mathbb{Z}^2)^*} \Phi_{\beta}^t(a, x) $, where $\Phi_{\beta}^t : (\mathbb{Z}^2)^* \times \Omega \rightarrow \mathbb{R}$ is such that
\begin{enumerate}
\item there is a function $\Phi_{v, \beta}^t : [0,2\pi)^v \rightarrow \mathbb{R}$ with $\Phi_{\beta}^t(v, x) = \Phi_{v, \beta}^t(x)$ for all $x$. This is trivially satisfied since we can set $\Phi_{v, \beta}^t(x) = \frac{1}{2} \sum_{i \sim j \in v} \Phi_{\beta}^t(x_i,x_j)$. Our symmetric interaction $\Phi_{\beta}^t(x_i)$ can be written as a sum over adjacent sites $\Phi_{\beta}^t(x_i,x_j)$.
\item Translation and reflection invariance w.r.t. the horizontal and vertical plane at $1/2$ are also trivially satisfied, since our nearest neighbour potential is symmetric.
\item $\Phi_{v, \beta}^t$ is also continuous and we set $m \equiv \inf \Phi_{v, \beta}^t$.
\end{enumerate}
We introduce in the sense of Georgii a low-energy site percolation on the dual lattice as follows. Let $\delta > 0$ be fixed then for each configuration $x \in \Omega$ a subgraph $G_{\delta}(x)=(V_{\delta}(x),E_{\delta}(x))$ of $(\mathbb{Z}^2)^*$ consists of the vertex set
\begin{equation*}
V_{\delta}(x) = \lbrace a \in (\mathbb{Z}^2)^* : \Phi_{v, \beta}^t(a,x) \leq m + \delta \rbrace
\end{equation*}
and edges connecting adjacent sites in $V_{\delta}(x)$. Let $C_{\delta}$ denote the cluster built on sites of $G_{\delta}(x)$. $\lbrace |C_{\delta}| = \infty \rbrace$ denotes the event that there is an infinite cluster $C_{\delta}$ built on low-energy configurations. The graph $G_{\delta}$ describes the corresponding low-energy clusters on the dual lattice.
Now we can prove the following.
\begin{propo}
For $\beta(t)$ large enough there is a time interval $(t_0,t_1)$ such that $|\mathcal{G}_{\beta}(\textbf{H}_{\beta}^t(\cdot,y^{spec}),\nu_0 )| \geq 2$.
\end{propo}
\textbf{Proof:}\\
The proof of this proposition will consist of proving two lemmas. First we prove the existence of a unique low-energy cluster and afterwards we deduce from that the spontanous breakdown of discrete symmetry and the existence of at least two distinct infinite-volume Gibbs measures. We will basically apply the proofs of Georgii's paper, \cite{Geo81}, to our simpler case. We consider finite-volume Gibbs distributions $\nu_{\beta, \Lambda}^\textbf{t}$, corresponding to the joint system conditioned at time $t$ with periodic boundary conditions, that is
\begin{equation}
\nu_{\beta, \Lambda}^\textbf{t}(dx) = \frac{e^{-\textbf{H}^t_{\beta}(x, y^{spec})}}{Z^{y^{spec}}_{\beta, \Lambda}} \nu_0^{\otimes \Lambda}(dx),
\end{equation}
with $\Lambda = \Lambda_n$.
\begin{Lemma}
Let $(\mathbb{Z}^2)^*$ indicate the dual lattice of the square lattice. Let $\delta > 0$ and $a \in (\mathbb{Z}^2)^*$ be fixed. Then for sufficiently large $\beta(t)$ there is a time interval $(t_0,t_1)$ such that there exists a.s. a unique infinite cluster $C_{\delta}$ for at least one translation invariant Gibbs measure $\nu_{\beta}^{\textbf{t}}$, i.e.
\begin{equation}
\nu_{\beta}^{\textbf{t}}(\lbrace \exists ! \text{ } |C_{\delta}| = \infty \rbrace) > 0.
\end{equation}
\end{Lemma}
We will sketch the proof of this lemma. The main ingredient which does the work is the following estimate.
\begin{Lemma}
Let for each $a \in \Lambda^* \subset (\mathbb{Z}^2)^*$ the function $f_a$, $f_a : [0,2\pi)^{v} \rightarrow [0,\infty)$ be given. $f_a$ is invariant under reflections of $v$. Then for all $\beta$
\begin{equation}
\int \prod_{a \in \Lambda^*} f_a(x_a) \nu_{\beta, \Lambda}^\textbf{t}(dx) \leq \prod_{a \in \Lambda^*} \biggl [ \prod_{b \in \Lambda^*} f_a(x_b) \nu_{\beta, \Lambda}^\textbf{t}(dx) \biggr ]^{1/|\Lambda|}. \label{chessboard}
\end{equation}
\end{Lemma}
This estimate is a consequence of the chessboard estimates which follow from the Reflection Positivity of $\nu_{\beta, \Lambda}^\textbf{t}$ w.r.t. reflections in the pairs of hyperplanes $R_1$ and $R_2$, where
\begin{equation*}
R_k = \lbrace z \in \mathbb{R}^2 : z_k = 0 \text{ or } n \rbrace.
\end{equation*}
To see the Reflection Positivity, Georgii points out that given the spin configuration in $\Lambda \cap R_k$, the spins in the remaining two parts are conditionally independent and up to reflections identically distributed. This applies clearly to our measure $\nu_{\beta, \Lambda}^\textbf{t}$. Using this inequality as a main ingredient, Georgii manages to show that for a fixed $\delta > 0$ the probability that there exists an infinite low-energy cluster $C_{\delta}$ in the dual lattice including the origin, tends to 1 as $\beta$ is going to infinity. His argument on the uniqueness of the low-energy cluster is based on translation invariance of $\nu^{\textbf{t}}_{\beta}$, which is also trivially satisfied in our case.
\begin{flushright}
$\square$
\end{flushright}
Reflection Positivity of the measure and the estimate given above
provide now the following lemma.
\begin{Lemma}
There is a function $r: [0,\infty) \rightarrow [0,1]$ with $r(\beta) \rightarrow 0$ for $\beta \rightarrow \infty$ such that for all positive $\beta$ and $\nu^{\textbf{t}}_{\beta} \in \mathcal{G}_{0, \beta}(\Phi^t_{v,\beta},\nu_0 )$ and all finite $D \subset (\mathbb{Z}^2)^*$ the inequality
\begin{equation*}
\nu^{\textbf{t}}_{\beta}(V_{\delta}(\cdot) \cap D = \emptyset) \leq r(\beta)^{|D|}
\end{equation*}
holds.
\end{Lemma}
The lemma states that the probability that in any finite set $D$ of the dual lattice there are \textbf{no} points coming from the low-energy cluster vertices $V_{\delta}$ is going to 0 for large $\beta$.
From this lemma he can conclude the existence of an infinite cluster. In fact the function $r$ is given by $r(\beta) = 1 \wedge e^{-\beta c_{\delta}}$ for some constant $c_{\delta}$ depending on $\delta$. Uniqueness can be deduced now from FKG inequalities and the 0-1 law for tail events or Georgii's argument which is heavily based on translation invariance of the measure.

\medskip

In the following we want to prove that the existence of a unique infinite low-energy cluster implies the existence of Gibbs measures in $\mathcal{G}_{\beta,0}(\Phi_{v, \beta}^t,\nu_0)$ for which the discrete left-right symmetry is broken. We will prove the following statement which we borrow from \cite{Geo81} in our simpler case.
\begin{Lemma}
Suppose there is an $\delta > 0$ such that the set $\lbrace x \in ([0,2\pi)^v : \Phi_v(x) \leq m + \delta \rbrace$ splits into two disjoint sets $A^{ri}$ and $A^{le}$ satisfying the stability and symmetry condition. Let $\beta$ be large such that $\nu^{\textbf{t}}_{\beta}(\lbrace \exists! |C_{\delta}| = \infty \rbrace) > 0$ for some $\nu^{\textbf{t}}_{\beta} \in \mathcal{G}_{0,\beta}(\Phi_{v, \beta}^t,\nu_0) \subset \mathcal{G}_{\beta}(\Phi_{v, \beta}^t,\nu_0)$ then there are at least 2 distinct measures extremal Gibbs measures $\nu^{\textbf{t}, ri}_{\beta}$ and $\nu^{\textbf{t},le}_{\beta}$.
\end{Lemma}
\textbf{Proof:}\\
We basically adapt the proof in \cite{Geo81} for our case, where we have a left-right symmetry which we want to break, and 2 symmetry-related ground states.

\medskip

Let $\delta > 0$ and let us look at the set $\lbrace \Phi \leq m + \delta \rbrace \subset [0,2\pi)^v$. This is the set of all configurations which correspond to a raise of energy by the amount of $\delta$ away from the ground state. Clearly this set splits into two disjoint measurable sets $A^{ri}$ and $A^{le}$, since we have exactly two ground states $x^{ri}$ and $x^{le}$. This splitting is also \textbf{stable} in the sense of Georgii, because two elements can only coincide on an edge of $v$ if they come from one of the sets $A^{ri}$ or $A^{le}$ (spins in $A^{ri}$ point to the right and those in $A^{le}$ point to the left).

\medskip

Let $ r_i: [0,2\pi)^v \rightarrow [0,2\pi)^v$, for $i=1$ or $2$, be the reflections of $v$ with respect to the hyperplanes $\lbrace z= (z_1,z_2) : z_i= 1/2 \rbrace$. Let $x$ be a configuration on $[0,2\pi)^{\mathbb{Z}^2}$ and let $A \subset [0,2\pi)^v$. Georgii constructs a graph $G_A(x)=(V_A(x),E_A(x))$ in the following way. The vertices $V_A(x)$ are those sites for which
\begin{eqnarray}
x_a \in
\begin{cases}
& A \text{ if } a \equiv v\mod 2 \\
& r_1(A) \text{ if } a \equiv v +(1,0)\mod 2 \\
& r_2(A) \text{ if } a \equiv v +(0,1)\mod 2 \\
& r_1(r_2(A)) \text{ if } a \equiv v +(1,1)\mod 2. \\
\end{cases}
\end{eqnarray}
The set $r_1(A)$ (resp. $r_2(A)$) is the reflected set $A$ through the vertical (resp. horizontal) line at $1/2$. Similarly, $r_1(r_2(A))$ denotes the set $A$ after being reflected first horizontally and then vertically. For a graph built on $A^{ri}$, it means we collect $a$ for which
\begin{eqnarray}
x_a \in
\begin{cases}
& A^{ri} \text{ if } a \equiv v\mod 2 \text{ or } a \equiv v +(0,1)\mod 2 \\
& A^{le} \text{ if } a \equiv v +(1,0)\mod 2 \text{ or } a \equiv v +(1,1)\mod 2 \\
\end{cases}
\end{eqnarray}
and for $A^{le}$ the graph is the same up to an $r_1$-reflection.

\medskip

This implies that each cluster $C_{\delta}$ of the graph $G_{\delta}(x)$ is a cluster $C_{A^{ri}}$ of $G_{A^{ri}}(x)$ or $G_{A^{le}}(x)$. Therefore the event that there exists a unique infinite cluster splits into a disjoint union of events that this infinite cluster appears in either $A^{ri}$ or $A^{le}$, namely
\begin{equation*}
\lbrace \exists ! | C_{\delta} | = \infty \rbrace = \lbrace \exists ! | C_{A^{ri}} | = \infty \rbrace \sqcup \lbrace \exists ! | C_{A^{le}} | = \infty \rbrace
\end{equation*}
where $\sqcup$ denotes the disjoint union. The splitting is also symmetric because of the simple relation $r_1(A^{le}) = A^{ri}$ (we have defined $r_1$ as the reflection w.r.t. the horizontal line at $z_1 = 1/2$). Thus it follows that for $\nu^{\textbf{t}}_{\beta} \in \mathcal{G}_{0,\beta}(\Phi_{v, \beta}^t, \nu_0)$
\begin{equation}
\nu^{\textbf{t}}_{\beta} (\lbrace \exists ! | C_{A^{ri}} | = \infty \rbrace ) = \nu^{\textbf{t}}_{\beta} (\lbrace \exists ! | C_{A^{le}} | = \infty \rbrace ) \label{symmetrie}
\end{equation}
and in particular that $\nu^{\textbf{t}}_{\beta} (\lbrace \exists ! | C_{\delta} | = \infty \rbrace) = 2 \nu^{\textbf{t}}_{\beta} (\lbrace \exists ! | C_{A^{ri}} | = \infty \rbrace )$. Since we know from the previous lemma that for $\beta$ large enough $\nu^{\textbf{t}}_{\beta} (\lbrace \exists ! | C_{\delta} | = \infty \rbrace) > 0$ we have a fortiori
\begin{equation}
\nu^{\textbf{t}}_{\beta} (\lbrace \exists ! | C_{A^{ri}} | = \infty \rbrace ) > 0
\end{equation}
which allows us to build conditional probabilities
\begin{equation*}
\nu^{\textbf{t},ri}_{\beta} := \nu^{\textbf{t}}_{\beta} (\cdot \text{ } | \lbrace \exists ! | C_{A^{ri}} | = \infty \rbrace )
\end{equation*}
resp.
\begin{equation*}
\nu^{\textbf{t},le}_{\beta} := \nu^{\textbf{t}}_{\beta} (\cdot \text{ } | \lbrace \exists ! | C_{A^{le}} | = \infty \rbrace )
\end{equation*}
which are orthogonal since $\lbrace \exists ! | C_{A^{ri}} | = \infty \rbrace$ and $\lbrace \exists ! | C_{A^{le}} | = \infty \rbrace$ are disjoint. The Gibbsianness of $\nu^{\textbf{t},ri}_{\beta}$ and $\nu^{\textbf{t},le}_{\beta}$ follows from the fact that $\lbrace \exists ! | C_{A^{ri}} | = \infty \rbrace$ (resp. $\lbrace \exists ! | C_{A^{le}} | = \infty \rbrace$) belongs to the tail $\sigma$-field $\mathcal{F}_{\infty}$. This is because the sets $\lbrace \exists ! | C_{A^{ri}} | = \infty \rbrace$ (resp. $\lbrace \exists ! | C_{A^{le}} | = \infty \rbrace$) are invariant under translations $\theta_i$, $i \equiv 0 \mod 2$, where $\theta_i$ denotes the translation by $i$. We have proved that the discrete symmetry $r_1$ is broken since there are at least two distinct Gibbs measures $\nu^{\textbf{t},ri}_{\beta}$ and $\nu^{\textbf{t},le}_{\beta}$.
\begin{flushright}
$\square$
\end{flushright}

\textbf{Remark:} The above transition is one of the so-called spin-flop type.

\section{Comments and possible generalizations }

In this section we'd like to discuss what kind of generalizations of our
results will hold.
In our first class of results, the conservation of Gibbsianness, the
restriction to finite-range potentials could be weakened, and also one could
obtain results for $N$-component spins for general $N \geq 3$. For some developments
in this direction, using Dobrushin uniqueness techniques, see \cite{KueOpo07}.

A more sensitive question is about the loss of the Gibbsian property.
Although we have given the proof for the loss of Gibbsianness for
the standard interaction in 2 dimensions,
where the initial Gibbs state is presumably unique at all temperatures,
the same arguments apply for other models, such as the nearest-neighbour
polynomial nonlinear models in 2 dimensions considered in \cite{vEnShl02}.

An extension to the three-dimensional lattice is also immediate. In fact,
because there is long-range order for any strength of the alternating
magnetic field including zero, the non-Gibbsianness holds for all times
larger than a certain $t_0$ in that case. Also the statement is true for
any initial translation-invariant (pure or not) Gibbs measure, of which
there are known to be infinitely many \cite{FroSimSpe76, BalOCa99}.

Increasing the spin-dimension, and considering $N$-component spins
is less immediate. In $d=2$, our proof breaks down. In $d=3$ however, a
ferromagnet in a small alternating field displays continuous symmetry-breaking
in the plane perpendicular to the field \cite{AizLieSeiSolYng04}. A slight modification of our argument then will lead to
non-Gibbsian behaviour.

A more serious limitation in our proof is that we used Reflection Positivity
in our proof of the phase transition. This restricts the initial interactions
quite severely (it should be a C-potential (generalized nearest-neighbor potential), or a pair interaction of very specific form,
and the dynamics needs to be an infinite-temperature one.
To get rid of this limitation, one might hope
to apply a Pirogov-Sinai type of approach. Dobrushin and Zahradnik
\cite{DobZah86, Zah00} have obtained Pirogov-Sinai results for continuous
spins. However, their
conditions (quadratic interactions with general single-site potentials
having Gaussian minima) do not directly apply. One might hope that their
ideas could be modified to include our set-up; however, till now this
has not been done. This seems a technically non-trivial question for future
research.
\medskip

\textit{Acknowledgements:} We thank Christof K\"ulske and Alex Opoku for many discussions and for making the results of \cite{KueOpo07} available to us. Also we thank Roberto Fern\'andez for a helpful discussion.



\begin{thebibliography}{}


\bibitem[AizLieSeiSolYng04]{AizLieSeiSolYng04}
M.Aizenman, E.H.Lieb, R.Seiringer, J.P.Solovej and J. Yngvason.\\
\textit{Bose Einstein quantum phase transition in an optical lattice model}\\
Phys. Rev. A 70, 023612, 2004.
\\
\bibitem[BalOCa99]{BalOCa99}
T. Balaban and M. O'Carroll \\
\textit{Low-temperature properties for Correlation Functions in Classical
N-Vector Spin models} \\
Comm. Math. Phys. 199, 493--520, 1999.
\\

\bibitem[BriFonLan77]{BriFonLan77}
J. Bricmont, J.R.Fontaine and L.J. Landau \\
\textit{On the uniqueness of the equilibrium state for plane rotators} \\
Comm. Math. Phys. 56, 281--290, 1977.
\\
\bibitem[CatRoeZes96]{CatRoeZes96}
P. Cattiaux, S.Roelly, H. Zessin \\
\textit{Une approche Gibbsienne des diffusions Browniennes infinie-dimensionelles} \\
Prob. Th. Rel. Fields 104, 147--179, 1996.
\\
\bibitem[Deu87]{Deu87}
J.D. Deuschel \\
\textit{Infinite-dimensional diffusion process as Gibbs measures on} $C[0,1]^{\mathbb{Z}^d}$ \\
Prob. Th. Rel. Fields 76, 325--340, 1987.
\\
\bibitem[DobZah86]{DobZah86}
R.L. Dobrushin, M.Zahradnik \\
\textit{Phase diagrams for continuous-spin models: An extension of the Pirogov-Sinai theory}
\\
Math. Problems of Stat. Phys. and Dynamics, Reidel, 1--123, 1986.
\\
\bibitem[DerRoe05]{DerRoe05}
D. Dereudre , S. Roelly \\
\textit{Propagation of Gibbsianness for infinite-dimensional gradient Brownian diffusions} \\
J.Stat.Phys. 121, 511--551, 2005.
\\
\bibitem[vEntFerHolRed02]{vEntFerHolRed02}
A.C.D. van Enter, R. Fernandez, F. den Hollander, F.Redig \\
\textit{Possible loss and recovery of Gibbsianness during the stochastic evolution of Gibbs measures} \\
Comm. Math. Phys., 226, 101--130, 2002.
\\
\bibitem[vEnFerSok93]{vEnFerSok93}
A.C.D. van Enter, R. Fernandez, A.D.Sokal \\
\textit{ Regularity Properties and Pathologies of Position-space
 Renormalization-Group Transformations: Scope and Limitations of
Gibbsian Theory.} \\
J.Stat.Phys. 72, 879--1169, 1993.
\\

\bibitem[vEnShl02]{vEnShl02}
A.C.D. van Enter and S.B. Shlosman \\
\textit{First-order transitions for n-vector models in two and more dimensions: Rigorous proof}\\
Phys.Rev.Lett. 89, 285702, 2002.
\\

\bibitem[FerPfi97]{FerPfi97}
R.Fernandez, C.-E. Pfister \\
\textit{Global specifications and quasilocality of projections of Gibbs measures}\\
Ann. of Prob. 25-3, 1284--1315, 1997.
\\

\bibitem[FerPro07]{FerPro07}
R. Fernandez, A. Procacci \\
\textit{Cluster expansion for abstract polymer models. New bounds from an old approach}\\
Comm. Math. Phys. 274, 123--140, 2007.
\\

\bibitem[FroSimSpe76]{FroSimSpe76}
J. Fr\"ohlich, B.Simon and T. Spencer.
\textit{Infrared bounds, phase transitions and continuous symmetry breaking}\\
Comm. Math. Phys. 50, 79--95, 1976.
\\
\bibitem[Geo81]{Geo81}
H.-O. Georgii \\
\textit{Percolation of low energy clusters and discrete symmetry breaking in classical spin systems}\\
Comm.Math.Phys. 81, 455--473, 1981.
\\

\bibitem[Geo88]{Geo88}
H.O. Georgii \\
\textit{Gibbs measures and phase transitions} \\
Berlin, W. de Gruyter, 1988.
\\
\bibitem[GolKuiLebMae89]{GolKuiLebMae89}
S. Goldstein, R. Kuik, J.L. Lebowitz, C.Maes \\
\textit{From PCA's to Equilibrium Systems and Back} \\
Comm. Math. Phys. 125, 71--79, 1989.
\\
\bibitem[GuiZeg02]{GuiZeg02}
A. Guionnet, B. Zegarlinski \\
\textit{Lectures on Logarithmic Sobolev Inequalities}\\
Lecture notes in mathematics, 36, 2002.
\\
\bibitem[KonMin97]{KonMin97}
Y. Kondratiev, R. Minlos \\
\textit{One-particle Subspaces in the Stochastic XY Model
} \\
J. Stat. Phys. 87, 613--643, 1997.
\\

\bibitem[KotPre86]{KotPre86}
R. Kotecky, D.Preiss \\
\textit{Cluster expansions for abstract polymer models}\\
Comm.Math.Phys. 103, 491--498, 1986.
\\
\bibitem[KueOpo07]{KueOpo07}
C. K\"ulske and A.Opoku \\
\textit{The posterior metric and goodness of Gibbsianness of tranforms
of possibly continuous spin models} \\
Groningen preprint, 2007.
\\
\bibitem[KueRed06]{KueRed06}
C. K\"ulske, F.Redig \\
\textit{Loss without recovery of Gibbsianness during diffusion of continuous spins} \\
Prob.Theor. Rel. Fields 135, no. 3, 428--456, 2006.
\\
\bibitem[Koz74]{Koz74}
O.K. Kozlov \\
\textit{Gibbs description of a system of random variables}\\
Probl. Info. Trans. 10, 258--265, 1974.


\bibitem[LeNRed02]{LeNRed02}
A. Le Ny, F. Redig \\
\textit{Short time conservation of Gibbsianness under local stochastic evolutions} \\
J. Stat. Phys. 109, 1073--1090, 2002.
\\
\bibitem[MaeNet02]{MaeNet02}
C.Maes, K.Netocny \\
\textit{Spacetime expansions for weakly coupled interacting systems}\\
J. Phys. A: Math. Gen 35, 3053--3077, 2002.
\\

\bibitem[MinRoeZes00]{MinRoeZes00}
R.A.Minlos, S.Roelly, H.Zessin \\
\textit{Gibbs states on space-time} \\
Pot. Analysis 13, 367--408, 2000.
\\
\bibitem[MinVerZag00]{MinVerZag00}
R.A. Minlos, A. Verbeure, V. Zagrebnov \\
\textit{A quantum crystal model in the light mass limit: Gibbs states}\\
Rev. Math. Phys. Vol. 12-7, 981--1032, 2000.
\\
\bibitem[OliPet06]{OliPet06}
M.J. de Oliveira and A.Petri \\
\textit{Temperature of non-equilibrium lattice systems}\\
International Journal of Modern Physics C 17, 1703--1715, 2006.
\\
\bibitem[PryBru95]{PryBru95}
J.M. Pryce, A.D. Bruce\\
\textit{Statistical mechanics of image restoration}\\
J. Phys. A 3, 511--532, 1995.
\\


\bibitem[Ros97]{Ros97}
S.Rosenberg\\
\textit{The Laplacian on a Riemannian Manifold: An Introduction to Analysis on Manifolds}\\
Cambridge University Press, 1997.
\\

\bibitem[SaiNis02]{SaiNis02}
Y.Saika, H. Nishimori \\
\textit{Statistical mechanics of image restoration by the plane rotator model}\\
J. Phys. Soc. Japan 71 (4), 1052--1058 APR, 2002.
\\

\bibitem[Sul73]{Sul73}
W.G. Sullivan \\
\textit{Potentials for almost Markovian random fields}\\
Comm. Math. Phys. 33, 61--74, 1973.
\\

\bibitem[Tan02]{Tan02}
K. Tanaka \\
\textit{Statistical mechanical approach to image processing}\\
J. Phys. A 37, R81--R150, 2002.
\\

\bibitem[You89]{You89}
L.Younes\\
\textit{Parametric Inference for Imperfectly Observed Gibbsian Fields}\\
Prob.Th. and Rel.Fields 82, 625--645, 1989.
\\

\bibitem[Zah00]{Zah00}
M. Zahradnik \\
\textit{Contour methods and Pirogov-Sinai theory for continuous spin lattice models} \\
Amer. Math. Soc. Transl. (2) 198, 2000.
\\



\end{thebibliography}
\end{document}